\documentclass[linenumbers]{aastex631}
\usepackage{epsfig,amsmath,amsfonts,amssymb,graphicx,subfigure,enumitem,color}
\usepackage{amssymb}
\usepackage{amsmath}
\usepackage[varg]{txfonts}
\usepackage{natbib}
\usepackage{url}             
\usepackage{graphicx}
\usepackage{float}
\newcommand{\beq}{\begin{equation}}
\newcommand{\eeq}{\end{equation}}
\newcommand{\beqa}{\begin{eqnarray}}
\newcommand{\eeqa}{\end{eqnarray}}
\newcommand{\beqann}{\begin{eqnarray*}}
\newcommand{\eeqann}{\end{eqnarray*}}
\bibpunct{(}{)}{;}{a}{}{,} 

\usepackage{textcomp}
\newcommand{\foo}{\rlap{+}{\texttimes}}
\shorttitle{QPP in Kink Unstable Jet}
\shortauthors{Mishra et al.}

\begin{document}
\nolinenumbers
\title{Origin of Quasi-Periodic Pulsation at the Base of Kink Unstable Jet}
\author[0000-0003-2129-5728]{Sudheer K.~Mishra}
\affil{Indian Institute of Astrophysics, Koramangala, Bangalore-560034, India.}

\author{Kartika Sangal}
\affil{Department of Physics, Indian Institute of Technology (BHU), Varanasi-221005, India}

\correspondingauthor{Pradeep Kayshap}
\author[0000-0002-1509-3970]{Pradeep Kayshap}
\email{virat.com@gmail.com}
\affil{VIT Bhopal, Kothari Kalan, Sehore, Madhya-Pradesh 466114, India}
\author[0000-0002-7208-8342]{Petr Jel\'{i}nek}

\affil{University of South Bohemia, Faculty of Science, Department of Physics\\
Brani\v sovsk\'a 1760, CZ -- 370 05 \v{C}esk\'e Bud\v{e}jovice, Czech Republic}

\author{A.K.~Srivastava}
\affil{Department of Physics, Indian Institute of Technology (BHU), Varanasi-221005, India}
\author{S.P.~Rajaguru}
\affil{Indian Institute of Astrophysics, Koramangala, Bangalore-560034, India.}
\begin{abstract}
We study a blowout jet that occurs at the west limb of the Sun on August 29$^{th}$, 2014 using high-resolution imaging/spectroscopic observations provided by SDO/AIA and IRIS. An inverse $\gamma$-shape flux-rope appears before the jet{--} morphological indication of the onset of kink instability. The twisted field lines of kink-unstable flux-rope reconnect at its bright knot and launch the blowout jet at $\approx$06:30:43 UT with an average speed of 234 km s$^{-1}$. Just after the launch, the northern leg of the flux rope erupts completely. The time-distance diagrams show multiple spikes or bright dots, which is the result of periodic fluctuations, i.e., quasi-periodic fluctuations (QPPs). The wavelet analysis confirms that QPPs have a dominant period of $\approx$ 03 minutes. IRIS spectra (Si~{\sc iv}, C~{\sc ii}, and Mg~{\sc ii}) may also indicate the occurrence of magnetic reconnection through existence of broad $\&$ complex profiles and bi-directional flows in the jet. Further, we have found that line broadening is periodic with a period of $\approx$ 03 minutes, and plasma upflow is always occurs when the line width is high, i.e., multiple reconnection may produce periodic line broadening. The EM curves also show the same period of $\approx$ 03 minutes in different temperature bins. The images and EM show that this jet’s spire is mainly cool (chromospheric/transition region) rather than hot (coronal) material. Further, line broadening, intensity, and EM curves have a period of $\approx$03 minutes, which strongly supports that multiple magnetic reconnection triggers QPPs in the blowout jet.

\end{abstract}


\section{Introduction}
Solar jets are an integral part of the solar atmosphere, and they are an important feature of the mass/energy cycle within the solar atmosphere. Solar jets occur everywhere in the solar atmosphere. The solar jets have been classified into different categories based on different criteria, namely, (1) based on morphology{--}standard jets $\&$ blowout jets \citep{2010ApJ...720..757M, 2013ApJ...769..134M}, (2) based on the region of solar atmosphere where they occur{--} active-region jets, coronal hole jets, quiet-Sun jets, network jets, umbral jets, penumbral jets, and polar jets \citep{2011A&A...534A..62S,2013ApJ...770L...3K, 2018A&A...616A..99K, 2014Sci...346A.315T, 2016ApJ...816...92T, 2017A&A...598A..11M, 2018NatAs...2..951S}, and (3) based on the filter in which they observed{--}e.g., H-$\alpha$ jet, extreme ultraviolet (EUV) jets, ultraviolet jet (UV) jets, X-ray jets, and radio jets (e.g., \citealt{1992PASJ...44L.173S, 1997Natur.386..811I, 2007Sci...318.1591S, 2015Natur.523..437S,2015MNRAS.451.1117F, 2017ApJ...851...67S,2014A&A...567A..11Z, 2017ApJ...841...27N, 2019ApJ...870..113Z}). Solar surges are another category of solar jet-like features, and they are made of mainly cool plasma. Therefore, cool lines/filters (e.g., H$_{\alpha}$, Ca~{\sc ii} H $\&$ K line, Mg~{\sc ii} h $\&$ k lines, Si~{\sc iv}, IRIS/SJI~1400~{\AA}, AIA~304~{\AA}, and many more) use to observe the solar surges \citep[e.g.,][]{2000SoPh..196...79S, 2005A&A...444..265T, 2021MNRAS.505.5311K}. However, the surges can heat the plasma up to transition region/coronal temperature and then rapidly cools down to chromospheric temperatures \citep{2016ApJ...822...18N,2017ApJ...850..153N, 2018ApJ...858....8N,2021SoPh..296...84D}. In addition, \citet{2015ApJ...801..124Y} have added one more category to the solar jets, i.e., dark jets{--} no intensity enhancement, but the signature exists in the Doppler velocity.\\

The energy required to power the solar jets comes from the complex magnetic field through the most widely occurring process (i.e., magnetic reconnection) in the solar atmosphere. These solar jets are distributed all over the solar atmosphere as magnetic reconnection can occur anywhere within the solar atmosphere when suitable physical conditions are pronounced. Widely, the bipolar magnetic field reconnects with the pre-existing magnetic field, and produces the jets in the solar atmosphere (e.g., \citealt{1995Natur.375...42Y, 1996PASJ...48..353Y, 1996PASJ...48..123S, 2007Sci...318.1591S, 2008ApJ...683L..83N, 2016SSRv..201....1R}). \citet{2013ApJ...771...20M} have performed a detailed numerical simulation, based on the magnetic reconnection between an emerging bipolar magnetic field with the pre-existing open coronal magnetic field, to understand the triggering mechanisms and dynamics of solar jets. Further, in another remarkable work, \cite{2015Natur.523..437S} have shown that mini-filament can also trigger the solar jets, i.e., the eruption of the mini-filament trigger the magnetic reconnection which ultimately produces the solar jet. In this continuation, \citet{2017Natur.544..452W} has proposed the universal magnetic breakout model to trigger the jet, i.e., magnetic reconnection takes place due to the mini-filament eruption. However, it must be noted that this magnetic breakout model was already known $\&$ widely used in the triggering of the large-scale eruptions, and \cite{2018ApJ...852...98W} have extended this model to the formation of solar jets. As per the magnetic breakout model, the magnetic reconnection takes place due to the mini-filament eruption, and it finally triggers the jet in the solar atmosphere. Hence, we can say that magnetic reconnection is an integral process of the formation of solar jets. \\

There are at least three different theoretical explanations for plasma acceleration from various numerical simulations. In the first type of acceleration mechanisms, the plasma is accelerated from the magnetic reconnection site by the slingshot effect along the newly reconnected magnetic field lines \citep{1996PASJ...48..353Y, 2008ApJ...683L..83N, 2008ApJ...673L.211M}. The released energy from the magnetic reconnection can be deposited through various ways, e.g., adiabatic compression, Joule heating, accelerated particles, and shocks. Hence, this released energy from the magnetic reconnection can heat the plasma impulsively, and the strong pressure and temperature gradient will develop there that can induce the evaporation flows \citep{2001ApJ...550.1051S, 2003ApJ...593L.133M, 2012ApJ...759...15M}. The evaporation flows is the second plasma acceleration mechanism for the solar jets, and it is induced by the magnetic reconnection. The speed in this mechanism (i.e., evaporation flow) of the jet plasma is much slower than the speed of plasma attained through the slingshot effect. When the twisted closed magnetic field lines reconnect with the untwisted open field lines then the twist will transfer to the newly reconnected magnetic field lines. And, the newly reconnected magnetic field lines will exhibit untwisting motions, which is the third type of acceleration mechanism induced by the magnetic reconnection. The models, which are based on this mechanism, are known as the untwisting models of the solar jets \citep[e.g.,][]{1986SoPh..103..299S, 1995SoPh..156..245S, 1996ApJ...464.1016C, 2004ApJ...610.1129J, 2010A&A...510L...1K, 2013ApJ...769L..21A, 2013ApJ...771...20M, 2014ApJ...789L..19F}. The helical/rotational motions (i.e., an important feature of the solar jets) can be explained through these untwisting models of the solar jets. In addition, it can be noted that the helical/rotational motion is a main feature of cool emissions of the solar jets \citep[e.g.,][]{1996ApJ...464.1016C, 2001A&A...379..324H, 2010A&A...510L...1K, 2013RAA....13..253H}. The blowout jet mainly emits at cooler temperature, and they always exhibit a strong rotation \citep[e.g.,][]{2010A&A...521A..49S, 2013ApJ...769..134M}.\\

The physical process used in the untwisting models (i.e., reconnection between the twisted magnetic field and open magnetic field lines) is not the only way to trigger the solar jets and their helical or rotating motions. Various types of magnetohydrodynamic (MHD) instabilities (e.g., Rayleigh-Taylor (RT), Kelvin-Helmholtz (KH), Ballooning mode, convective-driven instability, radiatively-driven instability, heating-driven thermal instabilities, tearing mode, kink mode, sausage mode, helical/torsional mode, and current-sheet mode) are important physical process trigger the wide varieties of the solar phenomena, i.e., from the small-scale event (e.g., solar jets) to large-scale eruptions \citep{2018ApJ...856...86M, 2019ApJ...874...57M, 2021ApJ...923...72M}. Kink instability is one of the prominent mechanisms to trigger the solar jets \citep[see review;][]{2016SSRv..201....1R}. The observational and theoretical studies suggest that the rotational/twisting motions of the solar jets are directly linked with the helical kink instability \citep{1986SoPh..103..299S, 2009ApJ...691...61P, 2019FrASS...6...44L, 2021A&A...649A.179Z}.\\

The embedded magnetic dipole is key feature of the numerical simulations based on the kink-instability. Generally, the MHD instabilities are associated with the embedded magnetic bipole in such numerical simulations, and the kink instability takes place when threshold in energy or helicity or twist exceeds some critical values \citep[e.g.,][]{2009ApJ...691...61P}. The kink instability may evolve in a flux rope if the azimuthal component of the magnetic field exceeds some critical threshold \citep{2004psci.book.....A}. The theoretical and observational studies suggest different threshold values of twist/helicity for the kink instability, namely, 2$\pi$ \citep{1958PhFl....1..265K}, 2.5$\pi$ (\cite{1981GApFD..17..297H}), 2.6$\pi$ \citep[e.g.,][]{2009ApJ...691...61P, 2015A&A...573A.130P},and 1.3 turns \citep{2019FrASS...6...44L}. Hence, we can say that the threshold value of twist to trigger kink-instability varies a lot, and it depends on the magnetic field conditions. Now, this kink instability forces a loss of the stability of the whole stable magnetic field configuration \citep[i.e., the embedded magnetic flux in the uniform magnetic field;][]{2009ApJ...691...61P}, and it leads to the magnetic reconnection that drives the helical solar jets \citep{2015A&A...573A.130P}. However, such magnetic field configuration is not the only possibility for helical jets. The high-resolution observations have shown the existence of twisted flux rope, and the reconnection within the twisted magnetic structure can also produce such helical jets \citep{2010ApJ...718..981R, 2011ApJ...735L..18L, 2013ApJ...770L...3K}.\\

The solar jets may also be the source of MHD waves \citep{2007Sci...318.1580C, 2012A&A...537A.124Z} and quasi-periodic pulsations \citep[e.g.,][]{2012A&A...542A..70M, 2014A&A...561A.134Z}. The quasi-periodic pulsations (QPPs) are a phenomenon frequently associated with solar flares, and these QPPs in solar flares occur over a vast range of periods, i.e, from a fraction of a second to several minutes \citep[e.g.,][]{2010PPCF...52l4009N, 2016SSRv..200...75N, 2020A&A...642A.195K}. The different periods of QPPs may be related to the different physical processes in the solar atmosphere. Recently, \citet{2021SSRv..217...66Z} have reviewed the observational and theoretical aspects of the QPPs in solar and stellar flares, and they suggest more than fifteen different mechanisms in support of the formation of QPPs in solar/stellar flares. All these physical mechanisms for QPPs are primarily associated with either MHD waves or quasi-periodic regimes of magnetic reconnection. As we know that QPPs are very common phenomena associated with solar/stellar flares while the observational detection of QPPs in solar jets is very rare. So far, there are only a few observational works that report the existence of the QPPs in the solar jets \citep[e.g.,][]{2012A&A...542A..70M, 2014A&A...561A.134Z}. Similar to the QPPs in solar flares, it is reported that multiple magnetic reconnection can trigger the QPPs in blowout jets also \citep{2012A&A...542A..70M}. Here, we mention that QPPs are not well studied in solar jets as there are only a few reports of QPPs in solar jets so far. And, on top of this fact, we further say that solar jet, kink-instability, and the formation of QPPs are not investigated as per the best of our knowledge.\\
 
In the present scientific work, we provide a systematic observational study of solar jet triggered due to the kink instability, and subsequent evolution of quasi-periodic pulsations. The observations and data analysis are discussed in section~\ref{sect:obs_data}. In section~\ref{sect:results} we presents the observational results related to the eruption of kink unstable jets and quasi-periodic pulsation. Finally, the discussion and conclusion are presented in Section~\ref{sect:discuss}. 

\section{Observations and Data Analysis} \label{sect:obs_data}

The Atmospheric Imaging Assembly (AIA) onboard Solar Dynamics Observatory (SDO) provides the full-disk images of the Sun in several filters (i.e., AIA~4500{\AA}, AIA~1600~{\AA}, AIA~1700~{\AA}, AIA~304~{\AA}, AIA~171~{\AA}, AIA~211~{\AA}, AIA~131~{\AA}, and AIA~94~{\AA}; \citealt{2012SoPh..275...17L}). Some filters capture the emission from the extreme ultra-violet (EUV)/UV region, and one filter of AIA captures the emission from the visible waveband. Therefore, these AIA filters capture the emission of the Sun from the top of the photosphere to the lower corona. The spatial resolution of the EUV waveband is 1.5$''$ with 0.6$''$ per pixel width and a cadence of 12 seconds (see \citealt{2012SoPh..275...17L} for more details). We use multi-wavelength imaging observation obtained from AIA/SDO for this present work. The jet was triggered around $\sim$06:28 UT on August 29${th}$, 2014 near the west limb of the Sun, and the jet event ended at $\approx$07:20 UT.\\

The AIA images are used to understand the temporal evolution of this jet event. In addition to the temporal evolution, the AIA images can also be used to perform the differential emission measures (DEM). The Differential Emission Measure (DEM) is a very useful parameter to understand the thermal nature of this jet event. To estimate the DEM, we have used the method developed by \citet{2012A&A...539A.146H}. This method uses regularized inversion technique to perform the DEM using the warm/hot EUV channel of AIA/SDO (i.e.,  94~{\AA}, 131~{\AA}, 171~{\AA}, 193~{\AA}, 211~{\AA}, and 335 {\AA}). It is an automated method that uses the zeroth-order regularized inversion to return the DEM as a function of temperature. This regularized method provide a positive solution for the extracted DEM. We divide the temperature range between log $T$=5.0{--}7.5 K, with 25 temperature bins and an interval of log $\Delta T$=0.1 K.\\

In addition to AIA imaging observations, we also use imaging (slit-jaw images(SJI)) and simultaneous spectroscopic observations obtained from the Interface Region Imaging Spectrograph (IRIS). The IRIS telescope captures the emission from the lower solar atmosphere (chromosphere and transition region) using a slit-jaw imager (SJI), and IRIS takes images in far ultraviolet (FUV: 1331.56{--}1358.40 {\AA} and 1390.00{--}1406.79 {\AA}) and near ultraviolet (NUV: 2782.56{--}2833.89 {\AA}) wavebands. IRIS does not provide full-disk images rather it provided small field-of-view (FOV) images of the Sun in some filters, namely, Mg~{\sc ii}~2796.0~{\AA} filter, Si~{\sc iv}~1400~{\AA} filter, C~{\sc ii}~1330~{\AA} filter, and more filters \citep[for more details see; ][]{2014SoPh..289.2733D}. The IRIS has observed this jet event only in the C~{\sc ii}~1330~{\AA} filter with the temporal cadence of 10 seconds, and the full field-of-view (FOV) of SJI is 119$''\times$119$''$.\\
The IRIS also provides spectroscopic observations, and it observes some of the prominent spectral lines of interface-region, such as Mg~{\sc ii} k~2796{\AA}, Mg~{\sc ii} h~2803~{\AA}, C~{\sc ii}~1334.53~{\AA}, C~{\sc ii}~1335.66~{\AA}, Si~{\sc iv} 1393.76~{\AA}, and Si~{\sc iv} 1402.77~{\AA}. IRIS has observed the spectra of this particular jet event with an 8-step coarse raster observation of temporal cadence of 9.6 seconds. Hence, each raster file of this observation takes $\sim$77 seconds (i.e., 8.0$\times$9.6 = 76.8 seconds).\\  

Lastly, we mention that we have also utilized wavelet analysis to diagnose the quasi-periodic behavior of various light curves extracted from this jet event. We adopt the method of \cite{1998BAMS...79...61T} for the wavelet analysis of the time series. We have applied the wavelet analysis on AIA~304~{\AA}, AIA~171~{\AA}, AIA~131~{\AA}, and AIA~211~{\AA} filter observations. We computed the wavelet power, global power, and 95$\%$ significance levels using the method developed by \cite{1998BAMS...79...61T}. 

\section{Observational Results} \label{sect:results}
We used high-resolution multi-wavelength imaging and spectroscopic data of AIA and IRIS to study an eruptive jet on August 29$^{th}$, 2014. This jet was situated near the west limb, and the jet was initiated around~06:28 UT. Here, we have described the whole jet event in upcoming subsections.  
 \begin{figure*}
  \includegraphics[trim = 0.0cm 0.0cm 3.0cm 0.0cm, scale=1.0]{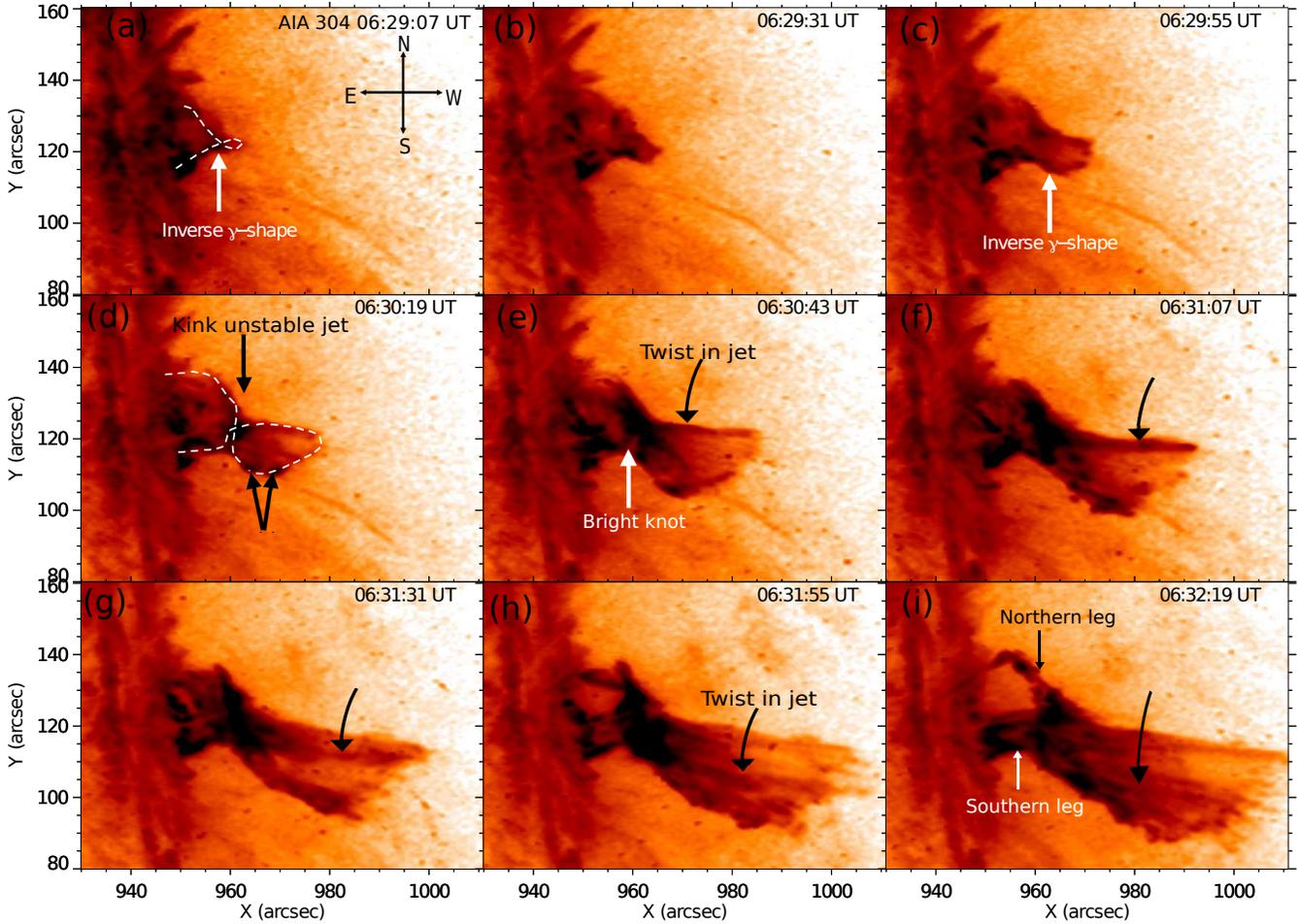}
\caption{The sequence of the images from SDO/AIA~304~{\AA} with reverse color contrast shows the onset of the kink instability in a blowout jet. The morphological sign of kink instability (inverse $\gamma$-shape), bright knot, propagation of brightening along the twisted field lines, and rotation of the plasma thread (i.e., twist in the jet) have been identified, and these observational findings are indicated by the different arrows. The first panel shows the direction to understand the dynamics in the jet's leg. The black and white arrow indicates the northern and southern legs of the eruptive jet (last panel).}
\label{fig:kink_unstable}
\end{figure*}
\subsection{Onset of the Kink-instability} \label{sect:kink}
We use a sequence of AIA 304~{\AA} images to understand the spatio-temporal evolution of kink instability in a blowout jet (cf., Figure~\ref{fig:kink_unstable}). The images of Figure~\ref{fig:kink_unstable} are plotted in the reverse intensity. The direction system is displayed in the panel (a) of Figure~\ref{fig:kink_unstable}. This blowout jet starts to lift at $\approx$06:28 UT from an active region NOAA 12146. 
At t = 06:29:07~UT, we see the inverse $\gamma$-shape structure outlined by a white-dashed line (panel a; Figure~\ref{fig:kink_unstable}), which is a manifestation of the writhing motion near the base of the jet. This magnetized structure (i.e., inverse $\gamma$-shape structure) is developing with time, i.e., it is expanding and rising (panels (b) and (c); Figure~\ref{fig:kink_unstable}; animation\_1.mp4, animation\_2.mp4). In the meantime, the inverse $\gamma$-shape structure performs an anticlockwise twist (see attached animation; animation$\_1$.mp4), and as a result, it may become kink unstable. The characteristic inverse $\gamma$-shape evolves due to the conversion of the initial twist into the writhe \citep{2010A&A...516A..49T, 2014PPCF...56f4012T}, and the inverse $\gamma$ shape-like structure is a morphological signature for the onset of the kink instability.\\

We see that the inverse $\gamma$-shape structure is lifting and expanding with time (panels (b) and (c); Figure~\ref{fig:kink_unstable}). Now, at time t = 06:30:19~UT, the inverse $\gamma$-shape structure has fully developed, and again we have outlined the fully developed structure by the white dashed line (panel (d); Figure~\ref{fig:kink_unstable}). We see little brightening around the knot (i.e., cross point) of the inverse $\gamma$-shape structure. 
In addition, we also see localized bright dots, and two bright dots are indicated by two black arrows (see panel d). The bright dots further propagate upward along the magnetic field lines as the jet develops with time. The brightening at the knot (as indicated by white arrows in the panel (e); Figure~\ref{fig:kink_unstable}) is becoming stronger and wider with time (see, panels (e) and (f); Figure~\ref{fig:kink_unstable}; animation\_1.mp4, animation\_2.mp4).\\

Meanwhile, the plasma is moving up along the magnetic field that is forming the main body of this jet. This jet is not well-collimated as the width of the jet is increasing with time (see, panels (d), (e), (f), (g), (h), and (i); Figure~\ref{fig:kink_unstable}; animation\_1.mp4, animation\_2.mp4). Hence, this jet is a typical blowout jet as per the morphological classification of the solar jets \citep[e.g.,][]{2010ApJ...720..757M, 2013ApJ...769..134M}. The jet is made of various plasma threads, and out of them, we have marked one plasma thread at 06:30:43~UT by a black arrow (panel (e); Figure~\ref{fig:kink_unstable}). This particular thread is located at the top edge of the jet this time (i.e., 06:30:43~UT), and further, we have traced this particular plasma thread at various other times (please see the black arrows from panel (e) to (l); Figure~\ref{fig:kink_unstable}). This particular thread follows anticlockwise motion with time, which suggests that the jet spire is rotating. In addition, we have also shown the evolution of the jet using IRIS~1330~{\AA} filter (cf., figure~\ref{fig:jet_evol_iris}). This animated figure describes the main features of the jet as per the IRIS animations (animation$\_$1.mp4, animation$\_$2.mp4). Similar to AIA~304~{\AA}, the IRIS/SJI~1330~{\AA} filter observations also show the rotating motion of the jet. Hence, finally, we can say that the rotating motion of the jet plasma is also visible in the animated figure~\ref{fig:jet_evol_iris}.
\subsection{Multi-wavelength imaging observations of solar jet} \label{sect:multi_jet}
\begin{figure*}
  \includegraphics[trim = 0.0cm 0.0cm 3.0cm 0.0cm, scale=1.0]{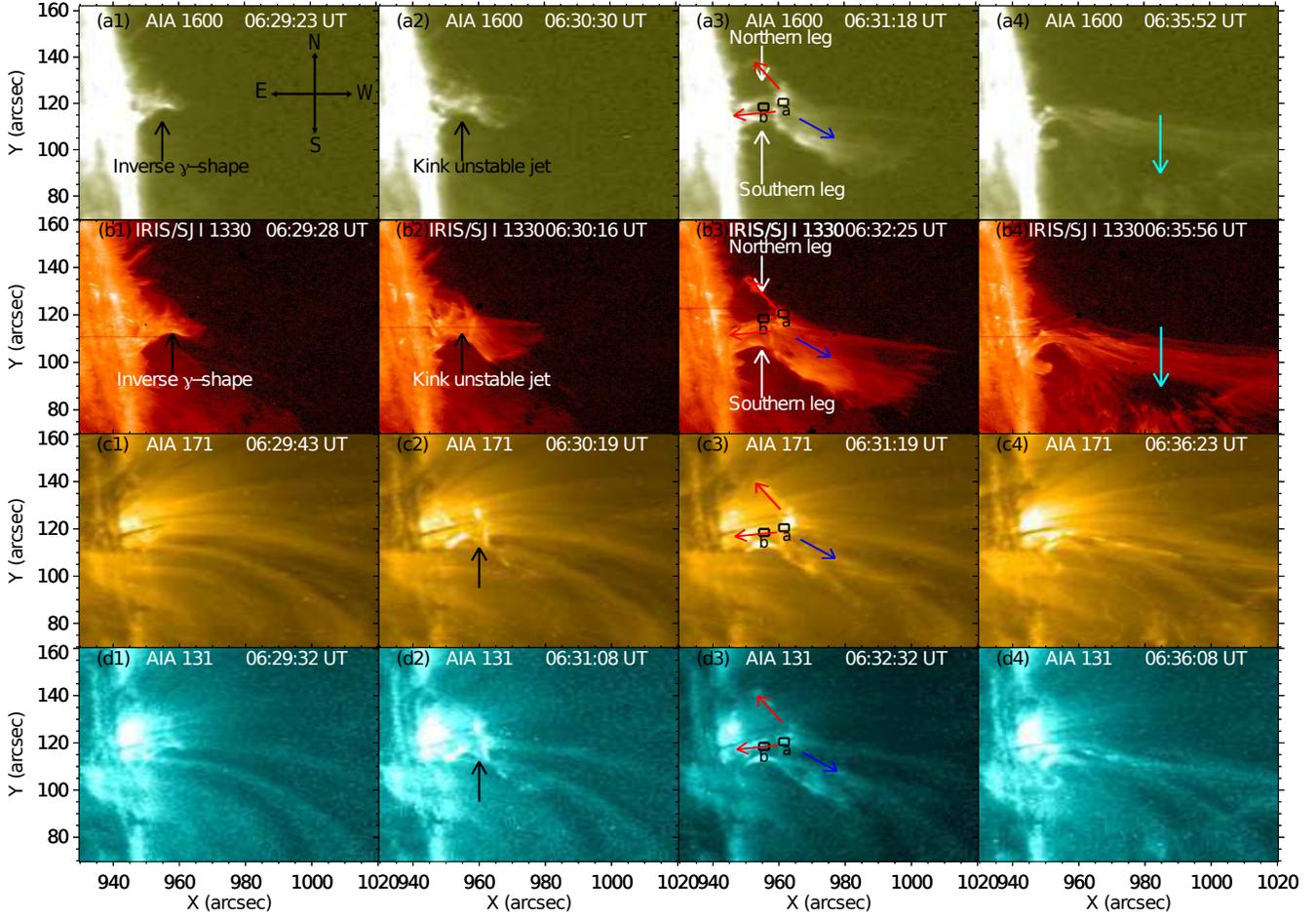}
\caption{The figures show the multi-wavelength view of the kink unstable blowout jet observed on August 29${th}$, 2014 by SDO/AIA and IRIS. It shows the morphological evolution of kink unstable blowout jet in the different layers of the solar atmosphere (i.e., AIA 1600 (top row), IRIS/SJI 1330 (second row), AIA 304 (third row), and AIA 131 (bottom row)). In this evolution, we have seen various important features of this blowout jet, namely, the development of inverse $\gamma$-shape in the cool filters (panels (a1) and (b1)), bright knot due to magnetic reconnection along with the triggering of blowout jet (panels (a2), (b2), (c2), and (d2)), bi-directional flows from bright know ((a3), (b3), (c3), and (d3)), and matured phase of jet with a cavity (panels (a4) and (b4)). Here, it should be noted that the jet is mainly visible in the cool filters, and does not emit much in hot filters. However, we see a compact bright structure in the hot filters (see (c2) and (d2)) which justifies the occurrence of internal magnetic reconnection in highly twisted magnetic field lines of inverse $\gamma$-shape near its apex. The northern and southern legs are indicated by white and black arrows in the panel (a3) and (b3). The two boxes 'a' and 'b' are shown in panels (a3), (b3), (c3), and (d3). We have used these boxes to investigate the total emission measure (EM) (see Figure~\ref{fig:dem_jet}(c) and(d)). The animation shows the (animation$\_$3.mp4; AIA 304, AIA 171, AIA 131, and AIA 94 {\AA}) triggering and the complete evolution of the kink unstable jet and their association with the multi-thermal plasma. The kink instability at the base of the jet, triggering of magnetic reconnection, the eruption of the jet, and associated dynamics are shown between 06:25 UT to 07:00 UT. The real-time duration of 12 s for this animation.}
\label{fig:jet_trigger}
\end{figure*}

In the previous subsection (i.e., section~\ref{sect:kink}), we have described the dynamics of the inverse $\gamma$-shape flux-rope and triggering of the blowout jet. In this subsection, we are investigating the dynamics of the blowout jet in cool (i.e., AIA~1600~{\AA}, AIA~304~{\AA}, and IRIS/SJI~1330~{\AA}) and warm/hot temperature filters (i.e., AIA~171~{\AA}, AIA~131~{\AA}, and AIA~94~{\AA}) to understand the multiwavelength nature of this jet. The distinction between hot and cool temperature filters is relative, and the classification between hot and cool temperatures may changes from case to case. Here, in the present study, we say that AIA~1600~{\AA}, AIA~304~{\AA}, and IRIS/SJI~1330~{\AA} are cool filters while AIA~171~{\AA}, AIA~131~{\AA}, and AIA~94~{\AA} are warm/hot filters. We know that this inverse $\gamma$-shape magnetic structure lifts up with time (see; Figure~\ref{fig:kink_unstable}). Here, in this subsection, we have discussed the multi-wavelength observations of the blowout jet. The Figure~\ref{fig:jet_trigger} shows the spatio-temporal evolution of the jet in AIA 1600~{\AA} (top row), IRIS/SJI~1330 {\AA} (second row), AIA~171~{\AA} (third row), and AIA~131~{\AA} wavebands. Similar to Figure~\ref{fig:kink_unstable}, we have again displayed the directions in the panel Figure~\ref{fig:jet_trigger}(a1). At t = 06:29:23~UT, the inverse $\gamma$-shape structure has significantly developed, and the top of this structure is well above the limb. This inverse $\gamma$-shape structure is clearly visible in cool temperature filters (panel (a1), (a2), (b1), and (b2);Figure~\ref{fig:jet_trigger}), but it is not visible in the hot temperature filters (panels (c1), (c2), (d1), and (d2); Figure~\ref{fig:jet_trigger}; animation\_3.mp4). However, we see that the brightened base of inverse $\gamma$-shaped structure in the hot temperature filters.\\

At the next time (t = 06:29:30~UT), we clearly see the jet in the cool temperature filter which is indicated by the black arrows (see, panels (a2) and (b2); Figure~\ref{fig:jet_trigger}). At this time, we see a compact brightening in the vicinity of knot of inverse $\gamma$-shape structure (indicated by black arrows in panels (c2) and (d2); Figure~\ref{fig:jet_trigger}) in the hot temperature filters (panels (c2) and (d2); Figure~\ref{fig:jet_trigger}). We do not see the spire of jet in the hot temperature filters as we have seen in cool filters. 
At the next time (t = 06:31:18~UT), the jet is fully developed with two legs rooted at the solar surface (panels (a3) and (b3); Figure~\ref{fig:jet_trigger}). The upper leg is termed as northern leg while the lower leg is termed as southern leg. And, these two legs are clearly indicated by the white and black arrows in the panels (a3) and (b3) of Figure~\ref{fig:jet_trigger}, respectively. Interestingly, we do see the signature of bi-directional flows, i.e., plasma falls along both legs below the bright knot (see red arrows in (a3), (b3), (c3), and (d3) panels; Figure~\ref{fig:jet_trigger}; animation\_3.mp4) while the plasma flows up above the bright knot (see blue arrows in (a3), (b3), (c3), and (d3) panels; Figure~\ref{fig:jet_trigger}). This bi-directional flow is also clearly visible in the animation$\_$1.mp4, animation\_2.mp4, animation\_3.mp4. This up-and-down flow of the plasma builds the main body of this blowout jet. Also, the bi-directional flows of the jet are visible in the cool filters. However, we only see some faint signatures of the blowout jet in the hot temperature filters (panels (c3) and (d3); Figure~\ref{fig:jet_trigger}). At time t = 06:35:52~UT, the northern leg of inverse $\gamma$-shape flux-rope is completely disconnected from the solar limb while, the southern leg remains connected to the limb of the Sun (panels (a4), (b4), (c4), and (d4); Figure~\ref{fig:jet_trigger}). Now, the jet is fully developed at this time along southern leg. Interestingly, we see big cavity in the jet plasma that is indicated by cyan color arrows (panels (a4) and (b4); Figure~\ref{fig:jet_trigger}). It seems that some plasma from the main body of the jet has bifurcated, and the space between main body and bifurcated plasma of jet appears the black region (i.e., cavity).   
 \subsection{Time-distance analysis of blowout jet}\label{sect:td}
\begin{figure*}
 \hspace{-1 cm}
\includegraphics[scale=1.0,angle=0,width=20.0cm,height=20.0cm,keepaspectratio]{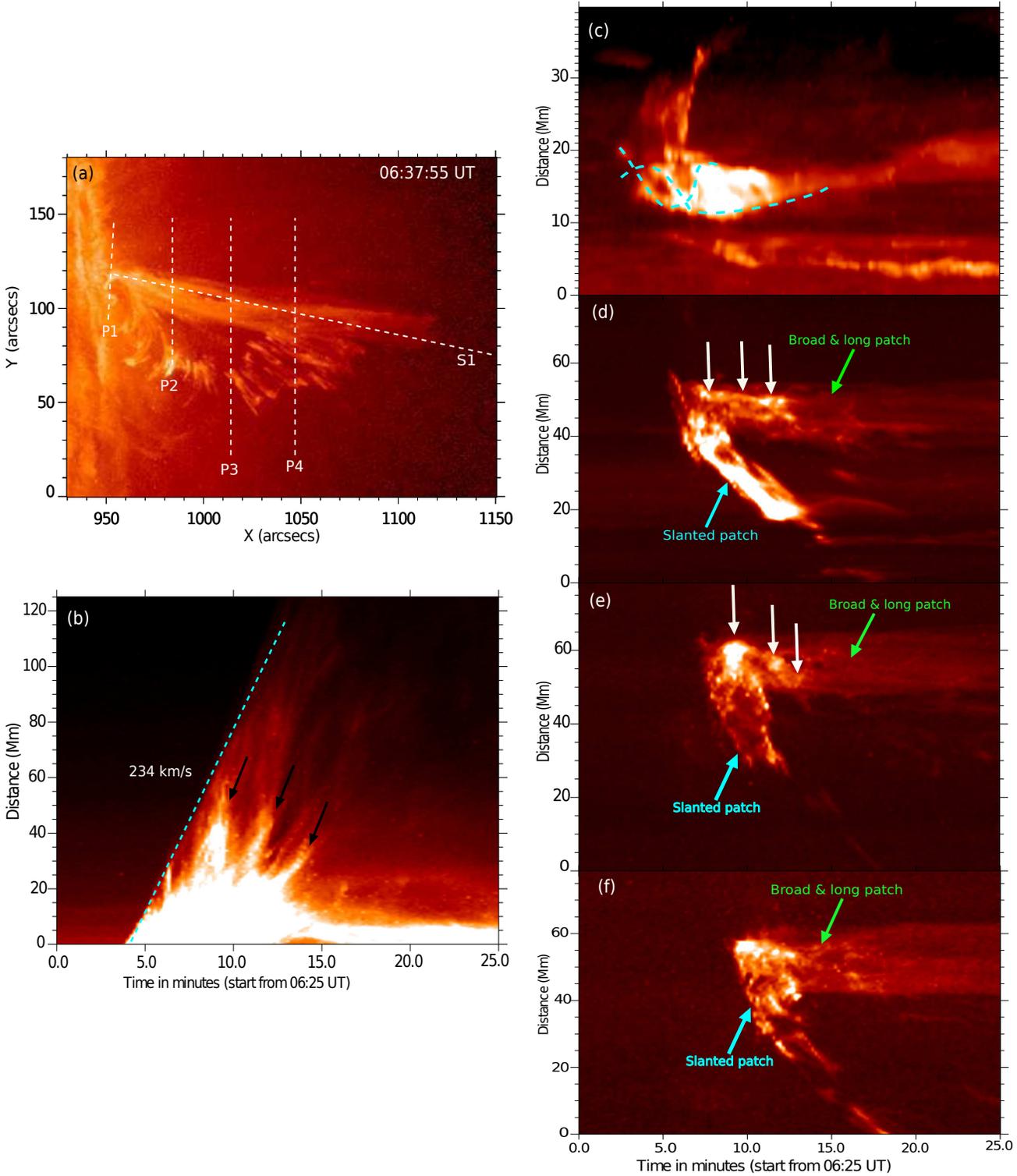}
\caption{The panel (a) shows the intensity image from AIA~304~{\AA} at t = 06:37:55~UT, i.e., from the decay phase of the jet. The over-plotted dashed lines are various slits along (i.e., S1) and across (i.e., P1, P2, P3, and P4) the jet, and they are used to produce the time-distance diagrams. The time-distance diagram along the slit S1 is shown in the panel (b), and we have drawn a line (i.e., dashed cyan line) along the ascending phase of the jet to estimate the speed of the blowout jet which is 234 km/s. We have also seen multiple spikes in this time-distance diagram which are indicated by the black arrows. In the right column, we have shown the time-distance diagrams along the slits P1 (panel (c)), P2 (panel (d)), P3 (panel (e)), and P4 (panel (f)). Here, we have seen the opposite motion of inverse $\gamma$-shape as indicated by the path drawn by the cyan dashed line in panel (c). Further, we have also seen the bright dots jet's body which is indicated by white arrows (panels (d) and (e)). Here, we see a slanted intensity patch (panels (d), (e), and (f)) in the last three panels of the right column which is occurring due to the fragmented plasma on a curved path from the main body of the blowout jet.} 
\label{fig:time_distance}
\end{figure*}

We have performed the time-distance estimation along and across (perpendicular) the blowout jet to understand the kinematics of this blowout jet (cf., Figure~\ref{fig:time_distance}). We displayed the late phase of the blowout jet along with one horizontal slit (S1) and four vertical slits (P1, P2, P3, and P4) in the panel (a) of Figure~\ref{fig:time_distance}. All of the slits are shown by white dashed lines (panel (a)). As we have already indicated that jet is mainly visible in the cool temperature filters, therefore, the time-distance analysis is performed using the cool temperature filter, i.e., AIA~304~{\AA}. The panel (b) of Figure~\ref{fig:time_distance} shows the time-distance diagram corresponding to the horizontal slit (i.e. slit S1). We have considered the width of 30 pixels around the horizontal slit (i.e., 15 pixels on each side of the slit) for the time-distance diagram shown in the panel (b) of Figure~\ref{fig:time_distance}. We have drawn a path (dashed cyan line) on the ascending motion of jet plasma, and it is found that the jet plasma is moving up with the speed of 234.0 km s$^{-1}$. Interestingly, on the close inspection, we noticed the multiple intensity enhancement at a regular intervals as indicated by the black arrows in the panel (b) of Figure~\ref{fig:time_distance}. Here, we have identified at-lest three peaks. \\

The right column of the Figure~\ref{fig:time_distance} shows four time-distance diagrams of the jet deduced from four different heights, i.e., along the slits P1 (panel (c)), P2 (panel (d)), P3 (panel (e)), and P4 (panel (f)). Again, we have used the same width of 30 pixels on both sides of all four slits (i.e., 15 pixels on both sides of the slit) in the production of time-distance diagrams. The panel (c) shows the time-distance diagram along the first vertical slit P1. We have already shown the presence of inverse $\gamma$-shape flux-rope, and the legs of this flux-rope show opposite motion before the jet eruption (see section~\ref{sect:multi_jet}). The slit P1 is located at the base of the jet, and covers both legs of the inverse $\gamma$-shape flux-rope. We observe two opposite motions of the plasma (white dotted curve on panel (c) of Figure~\ref{fig:time_distance}). This particular pattern is visible due to the opposite motion of the legs of the flux rope. 

The time-distance diagram as per the second slit P2 is displayed in the panel (d) of Figure~\ref{fig:time_distance}. In the first instance, we see two different intensity patches in the time-distance diagram, i.e., the first one is a long $\&$ broad patch (indicated by green arrow) while the second one is a narrow and slanted patch (indicated by cyan arrow). In the very initial phase, the blowout jet had a single body, while after some time the jet body was bifurcated into two parts with a cavity in between them as already explained in section~\ref{sect:multi_jet}. Even, in the reference image (panel (a); Figure~\ref{fig:time_distance}), one can see the long straight main body of the jet, and some plasma fragments are distributed on a curved path below the main body of the jet. The base of slit P2 lies on the curved path, and then the slit crosses some part of the cavity before covering the main body of the jet. Hence, the bottom narrow slanted intensity patch (indicated by cyan color arrow) in the time-distance diagram forms due to the plasma fragments along a curved path. While the long $\&$ broad patch (indicated by green arrow) is due to the dynamics of the main jet body. \\ 

The main body of the blowout jet contains small bright dots that are indicated by white arrows in the long straight intensity patch. The time-distance diagram along the slit S1 (i.e., panel (b); Figure~\ref{fig:time_distance}) has already shown the multiple intensity peaks. We observe that \textit{same intensity enhancement is visible as multiple bright dots on regular time-interval} in this time-distance diagram (panel (d) of Figure~\ref{fig:time_distance}) drawn for the across the jet (i.e., slit P2). Here, at least we have identified three bright dots, and it should be noted that these bright dots are aligned on the slanted path in the broad $\&$ long patch.\\

The time-distance diagram corresponding to slit P3 is displayed in the panel (e) of Figure~\ref{fig:time_distance}. Similar to the previous time-distance diagram, at this height, we also see the bright dots in the main body of the jet. Here, again these bright dots are indicated by the white arrows in the time-distance diagram corresponding to slit P3 (panel (e); Figure~\ref{fig:time_distance}), and these bright dots are on the slanted path in the long $\&$ broad patch (indicated by green arrow). The narrow slanted patch of intensity (indicated by cyan arrow) exists here as already seen in the panel (d) of Figure~\ref{fig:time_distance}. The last vertical slit (i.e., slit P4) is located very far away from the base of the jet. And, the panel (f) of Figure~\ref{fig:time_distance} shows the time-distance diagram corresponding to this slit. Again, we see both patches of intensity as already seen in the panels (d) and (e). Although, intensity in the slanted patch is weak in comparison to the panels (d) and (e) of Figure~\ref{fig:time_distance}. It is because that slit P4 is located near the top part of the jet. However, we see the multiple bright dots like the time-distance diagrams corresponding to the slits P2 (panel (d)) and P3 (panel (e) of Figure~\ref{fig:time_distance}).


\subsection{Thermal structure of blowout jet}\label{sect:dem}
\begin{figure*}
 \mbox{
  \includegraphics[trim = 0.0cm 0.0cm 0.0cm 0.0cm, scale=1.0]{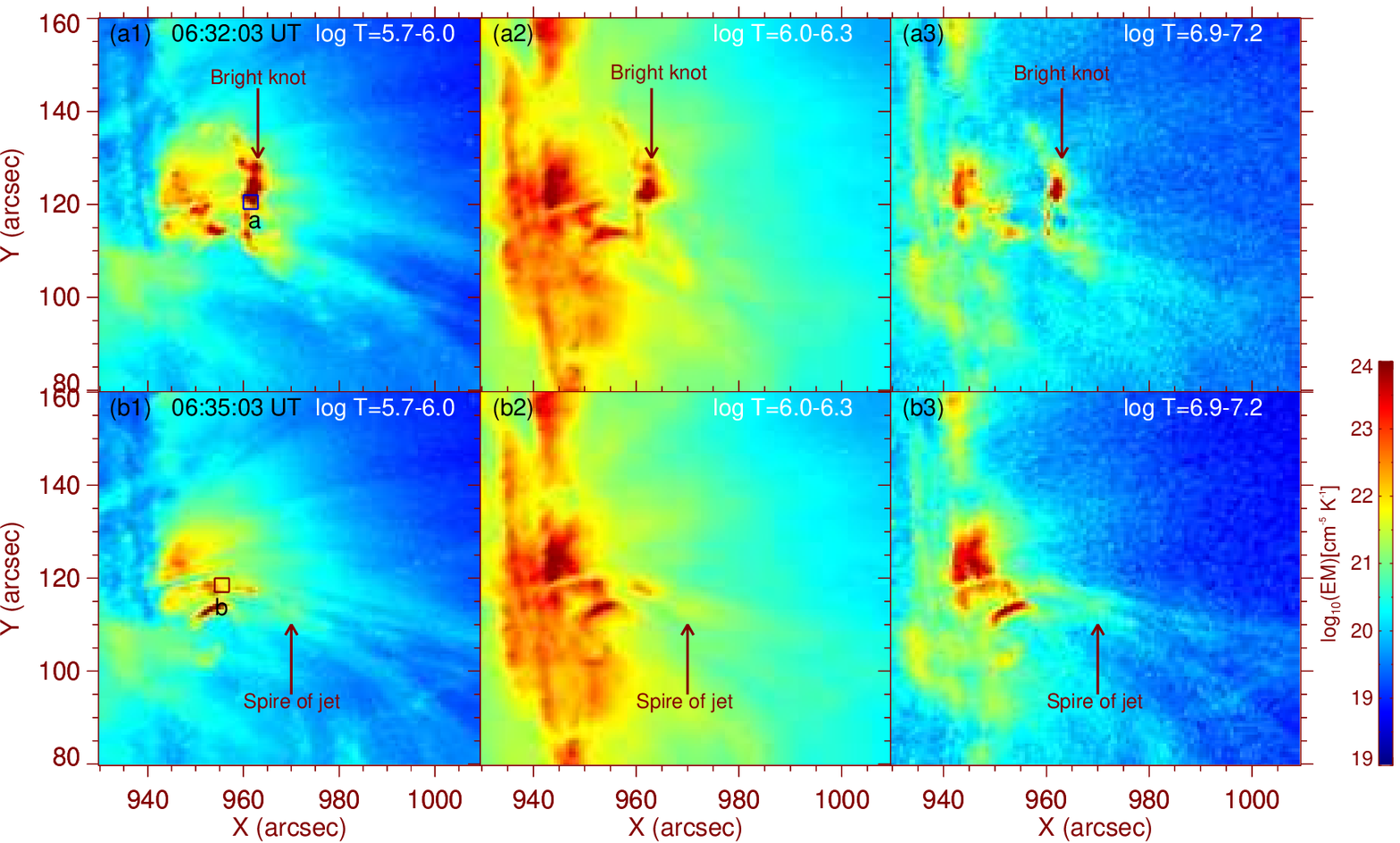}
  }
  \mbox{
  \includegraphics[trim = .0cm 0.0cm 0.0cm 0.0cm, scale=0.5]{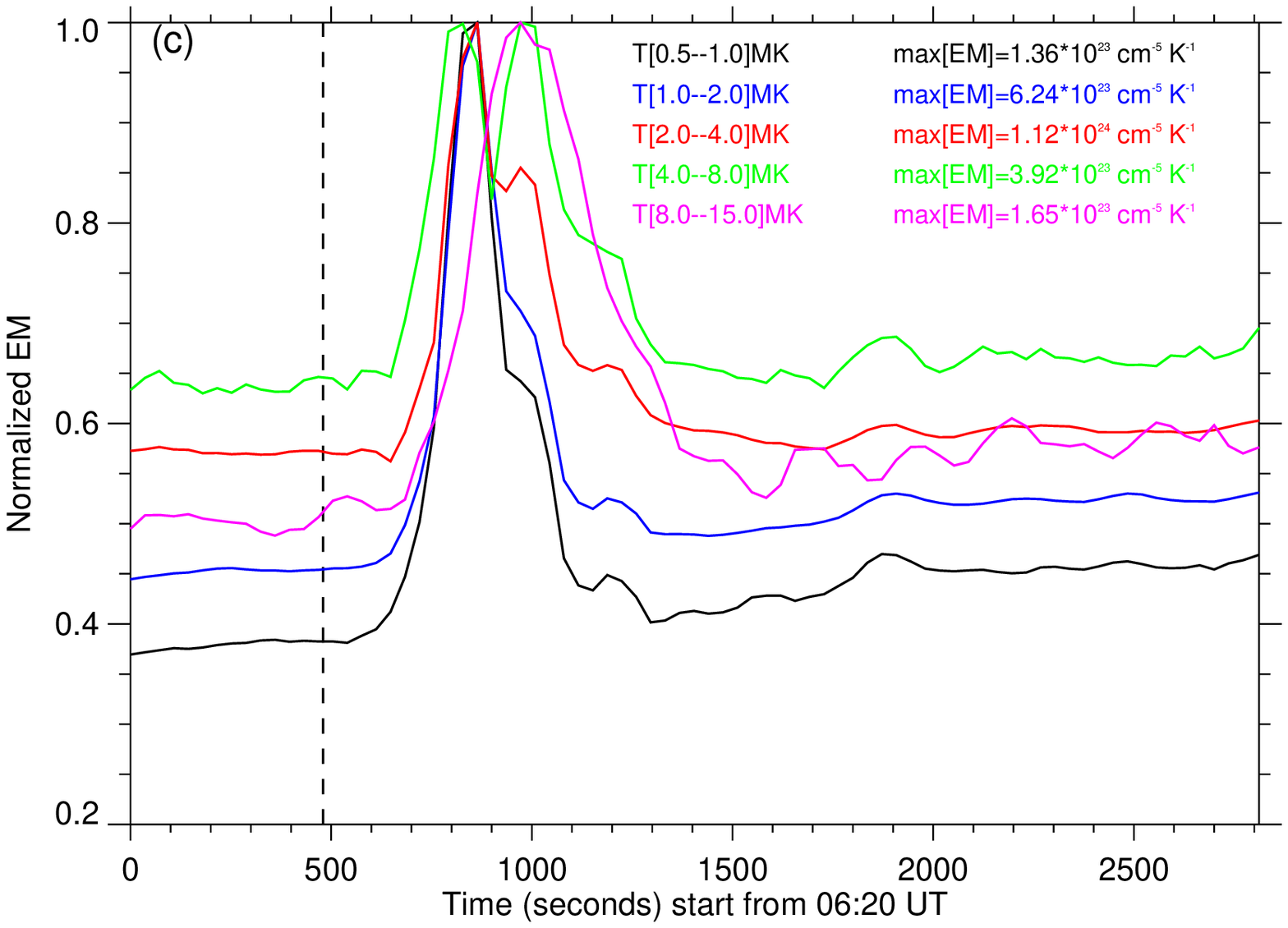}
    \includegraphics[trim = .0cm 0.0cm 0.0cm 0.0cm, scale=0.5]{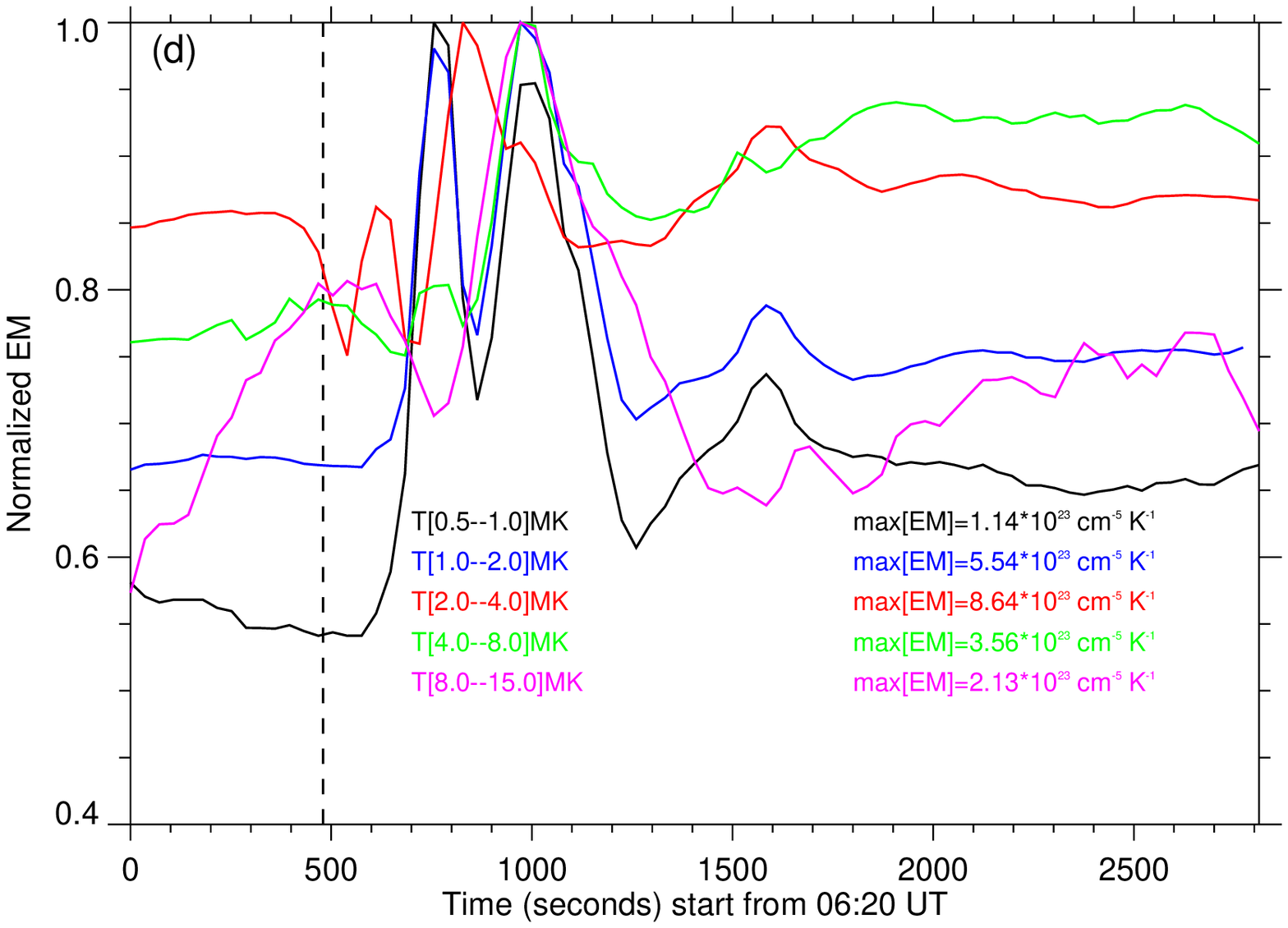}

  }
\caption{The panels (a1), (a2), and (a3) shows EM maps in three different temperature ranges (i.e.,  log T/K = 5.7{--}6.0, 6.0{--}6.3, and 6.9{--}7.2) during the time t = 06:32:03~UT when the magnetic reconnection takes place around the bright knot of inverse-$\gamma$ structure. The bright knot is indicated by the black arrows in all panels of the top row. Similarly, the middle row shows the same EM maps during the developed phase of this blowout jet (i.e., t = 06:35:03~UT). Here, we see very faint emission of the jet as indicated by black arrows in the middle row. Further, we have selected one box in the northern leg (i.e., box a) and another box in the southern leg (box b) to investigate the temporal variations of the EM curves in various temperature ranges. Various EM curves are shown in panels c (box a) and d (box b) of Figure~\ref{fig:dem_jet} by various colors. The color scheme and temperature ranges are mentioned in both panels. All EM curves from the box a show only one peak while all EM curves from box b show three to four peaks. The box (a) is located in the northern leg which erupts in the initial phase of the jet.}
\label{fig:dem_jet}
\end{figure*}

We perform the differential emission measure (DEM) analysis to understand the thermal nature of this blowout jet. We use regularized inversion code developed by \citep{2012A&A...539A.146H} to extract the DEM using hot AIA/SDO filters. We use six hot optically thin EUV filters (e.g., 94~{\AA}, 131~{\AA}, 171~{\AA}, 193~{\AA}, 211~{\AA}, and 335~{\AA}) of the AIA/SDO to estimate DEM coming from different temperatures bins. In the panels (a1), (a2), and (a3) of Figure~\ref{fig:dem_jet}, we have displayed three emission measure (EM) maps deduced from three different temperature ranges (i.e., log T/K = 5.7{--} 6.0, log T/K = 6.0{--} 6.3, and log T/K = 6.9{--} 7.2) during the onset of the kink-instability (i.e., t = $\sim$06:32:03~UT). Similarly, in the panels (b1), (b2), and (b3) of Figure~\ref{fig:dem_jet}, we have shown the same three EM maps during the developed phase of the blowout jet (i.e., t = $\sim$06:35:03~UT).\\

In the initial phase, EM maps show a significant emission around the bright knot (as defined previously in sections~\ref{sect:kink} and ~\ref{sect:multi_jet}) above the legs of inverse $\gamma$ flux-rope as indicated by the black arrows in (a1), (a2), and (a3) panels of the Figure~\ref{fig:dem_jet}. The emission in the bright knot region exists over a very wide range of the temperature, i.e., log T = 5.7 to 7.2 K. Hence, it justifies the presence of multi-thermal plasma at the knot of inverse $\gamma$-shape flux-rope.
After approximately 03 minutes, the emissions in the vicinity of the bright knot are significantly reduced (see all panels (b1), (b2), and (b3); Figure~\ref{fig:dem_jet}). Although we see little emission in the bottom region of the bright knot. And, a faint jet originating from this little emission area. The faint jet is indicated by the black arrows in the panels (b1), (b2), and (b3) of Figure~\ref{fig:dem_jet}.This EM study indicates that the spire of jet has very little hot emission. That is consistent with this jet showing up strongly in the cool filters, and being very faint in the hot filters.

We have selected a box (a) within the bright knot region (see, black rectangular box in panel (a1); Figure~\ref{fig:dem_jet}) to know the temporal behavior of DEM. We classify the entire DEM into five different temperature ranges (bins), namely, 0.5{--}1.0, 1.0{--}2.0, 2.0{--}4.0, 4.0{--}8.0, and 8.0{--}15 MK. Then, we estimated emission (EM = $\int_{T_{min}}^{T_{max}}{DEM(T) dT}$) in all above specified temperature bins. Through this approach, we got five different EM curves, and they are displayed by five different colors in the panel (c) of Figure~\ref{fig:dem_jet}. It is visible that all five curves show a dominant peak during the formation of the bright knot. The total emission in all five temperature ranges is highest during the formation of the bright knot.\\

We also extract the EM curves from another box (i.e., box (b)), which is situated inside the southern leg of the blowout jet (please see box (b) in panel (b1) of Figure~\ref{fig:dem_jet}). These EM curves are displayed in panel (d) of Figure~\ref{fig:dem_jet}. Interestingly, the EM curves from box (b) show periodic nature, unlike the nature of EM curves deduced from the box (a) (panel (c) of Figure~\ref{fig:dem_jet}). The box (a) is located in the northern leg of the blowout jet that erupts completely during the initial phase of the blowout jet. The plasma is completely swept away in the vicinity of the northern leg right after its eruption. Hence, we see only one dominant peak in EM curves extracted from the box (a). While, box (b) is situated near the base of the stable leg (i.e., southern leg) of this blowout jet. The periodic nature of EM curves extracted from the box (b) is present in all five temperature bins. We can easily locate at least three to four peaks in each EM curve, and it matches with the intensity light curve extracted from the different boxes from cool temperature filter AIA~304{\AA} (see, section~\ref{sect:QPP}). The vertical black dotted line in the lower panels (i.e., panels (c) and (d); Figure~\ref{fig:dem_jet}) indicates the jet event start time (t = $\sim$06:28~UT).

\subsection{Spectroscopic diagnosis of blowout jet}\label{sect:spectra}
In addition to the slit jaw images, IRIS has also captured the near ultraviolet (NUV) as well as far ultraviolet (FUV) spectrum of this jet event. 
\begin{figure*}
  \includegraphics[trim = 0.5cm 0.0cm 3.0cm 0.0cm, scale=1.0]{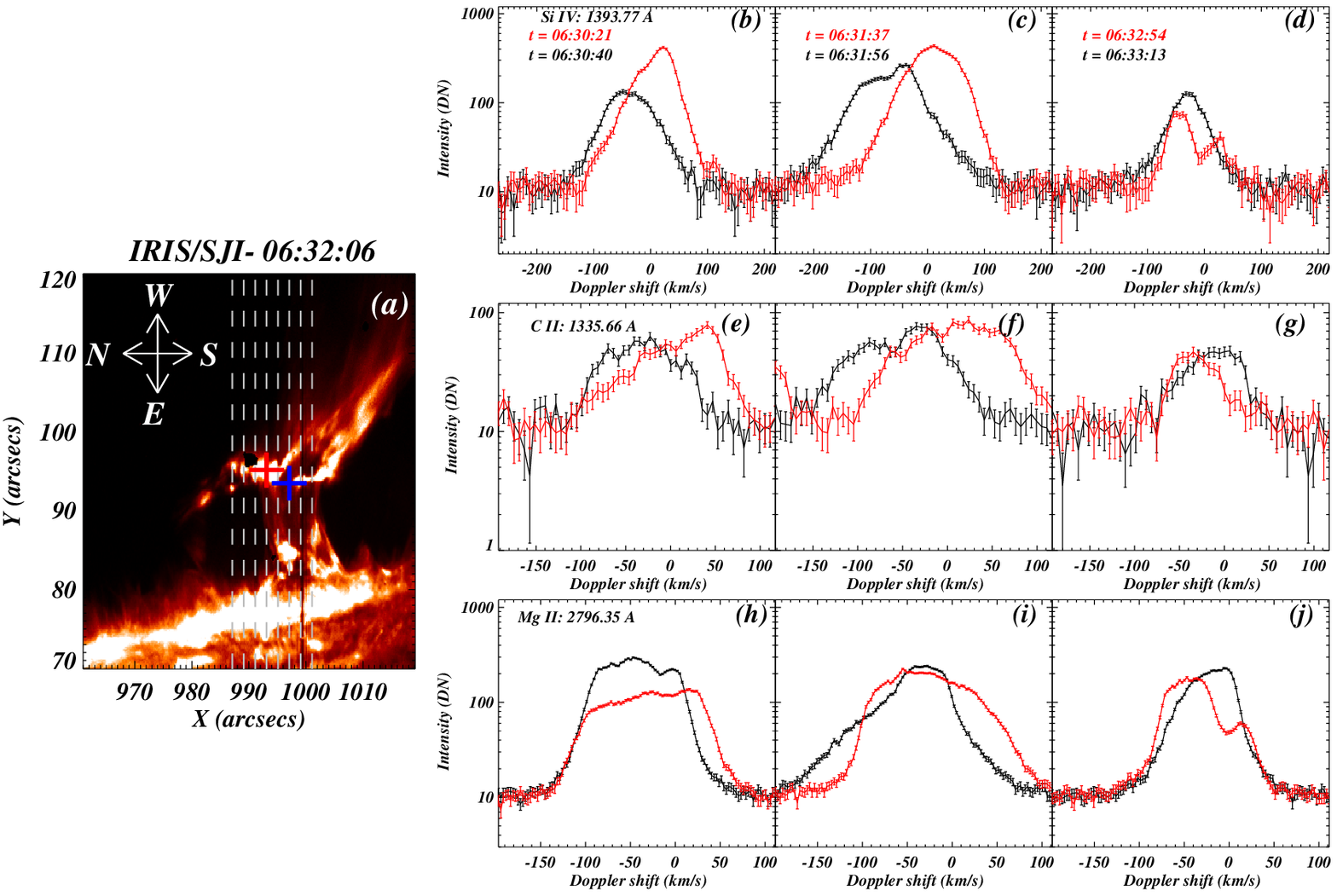}
\caption{IRIS/SJI 1330 {\AA} image depicts the blowout jet at time t = 06:32:06~UT along with 8-slit positions shown by white dashed lines in panel (a). IRIS has captured the spectra along these slits. Further, we have selected one location (i.e., red plus sign) in the northern leg while another location (blue plus sign) in the southern leg of the blowout jet. The panel (b), (c), and (d) show the spectral profiles of Si~{\sc iv} from both locations (i.e., the red curve from red plus location and black curve from blue plus location) at time t = 06:30~UT (panel b), 06:31~UT (panel (c)), and 06:32~UT (panel (d)). In the same fashion, we have shown C~{\sc ii} (panels (e), (f), and (g)) and Mg~{\sc ii} (panels h, i, and (j)) from both locations at the same three times. In general, all the spectral profiles from the red plus location are red-shifted (i.e., plasma downfall) while these profiles are blue-shifted (i.e., plasma upflows) from the blue plus location. In addition, all the spectral profiles are very complex profiles.} 
\label{fig:sji_spectra}
\end{figure*}
The panel (a) of Figure~\ref{fig:sji_spectra} shows the IRIS/SJI intensity map at a time of $\sim$06:32:06~UT. The IRIS has observed the event with a roll angle of 90$^{\circ}$. Therefore, this IRIS/SJI intensity map (i.e., panel (a)) of the Figure~\ref{fig:sji_spectra} is 90$^{\circ}$ rotated in comparison to other Figures (i.e., Figure~\ref{fig:kink_unstable}, ~\ref{fig:jet_trigger}, and other Figures) of this manuscript. Here, we mention that direction system is added in the panel (a) of Figure~\ref{fig:sji_spectra}. We have over-plotted all eight slit positions (i.e., eight vertical gray-dashed lines) on this IRIS/SJI intensity map (panel a). Then, we have chosen two locations to know the nature of Si~{\sc iv}, C~{\sc ii} and Mg~{\sc ii} profiles, namely, red plus sign (x = 993.12$"$ and y = 95.19$"$) and blue plus sign (x = 997.12$"$ and y = 93.52$"$). It should be noted that the selected locations (red and blue plus sign in panel a of Figure~\ref{fig:sji_spectra}) are located in the vicinity of the bright knot region, i.e., most probably in the magnetic-reconnection region.\\ 
In the imaging analysis, we have already shown the existence of the bi-directional flows in this jet event (see; section~\ref{sect:multi_jet}). And, as per the imaging analysis, the red plus sign lies in the down-flow region while the blue plus sign is in the up-flow region (see; Figure~\ref{fig:jet_trigger} and section~\ref{sect:multi_jet}). In Figure~\ref{fig:sji_spectra}, we have displayed Si~{\sc iv} 1393.77~{\AA} spectral profiles from red plus sign location by the red curve at three different times (i.e., 06:30:21~UT (panel b), 06:31:37~UT (panel c), and 06:32:54~UT (panel d)). Similarly, the Si~{\sc iv} 1393.77~{\AA} spectral profiles from blue plus locations are displayed by the black curve at three different times (i.e., t = 06:30:40~UT (panel (b)),~06:31:56~UT (panel (c)), and 06:33:13~UT (panel (d)). In the same way, we have shown the spectral profiles from red plus (red curve) and blue plus locations (black curve) of C~{\sc ii} 1335.66~{\AA} spectral profiles (see panels (e), (f), and (g)) and Mg~{\sc ii} 2796.35~{\AA} (see panels (h), (i), and (j)).

We have noticed that all black profiles (i.e., from Si~{\sc iv}, C~{\sc ii}, and Mg~{\sc ii} at all three times) are blue-shifted (upflows) while all the red profiles are red-shifted (down-flows). Hence, we can say that all three spectral lines (e.g., Si~{\sc iv}, C~{\sc ii}, and Mg~{\sc ii}) justify that red plus location is dominated by plasma downfall while the blue plus location is dominated by up flows, i.e., the spectra confirm the presence of bi-directional flows in the vicinity of the bright knot. Hence, finally, we can say that both (spectra and images) confirm the presence of bi-directional flows in the vicinity of the bright knot. In addition, we do see very broad profiles from Si~{\sc iv}, C~{\sc ii}, and Mg~{\sc ii} spectral lines during the initial/main phases of the blowout jet. Si~{\sc iv} 1393.77~{\AA} spectral line is an optically thin line, and normally, it is a single peak. However, in this blowout jet, the Si~{\sc iv} 1393.77~{\AA} spectral line is either double peak or highly asymmetric profile (panels (b), (c), and (d) of Figure~\ref{fig:sji_spectra}). On the other hand, C~{\sc ii} 1335.62~{\AA} and Mg~{\sc ii} 2796.35~{\AA} spectral lines are optically thick lines, and mostly, they appear as double peak profiles in the solar atmosphere. However, in this blowout jet, we see the very complex type of profiles from C~{\sc ii} 1335.62~{\AA} (see panels (e), (f), and (g)) and Mg~{\sc ii} 2796.35~{\AA} (see panels (h), (i), and (j)). Such type of complex profiles are reported in some small-scale energetic events \citep[e.g.,][]{2014Sci...346C.315P, 2018ApJ...857....5Y}. 
 \begin{figure*}
\mbox{
  \includegraphics[trim = 1.0cm 1.0cm 0.0cm 0.0cm, scale=1.2]{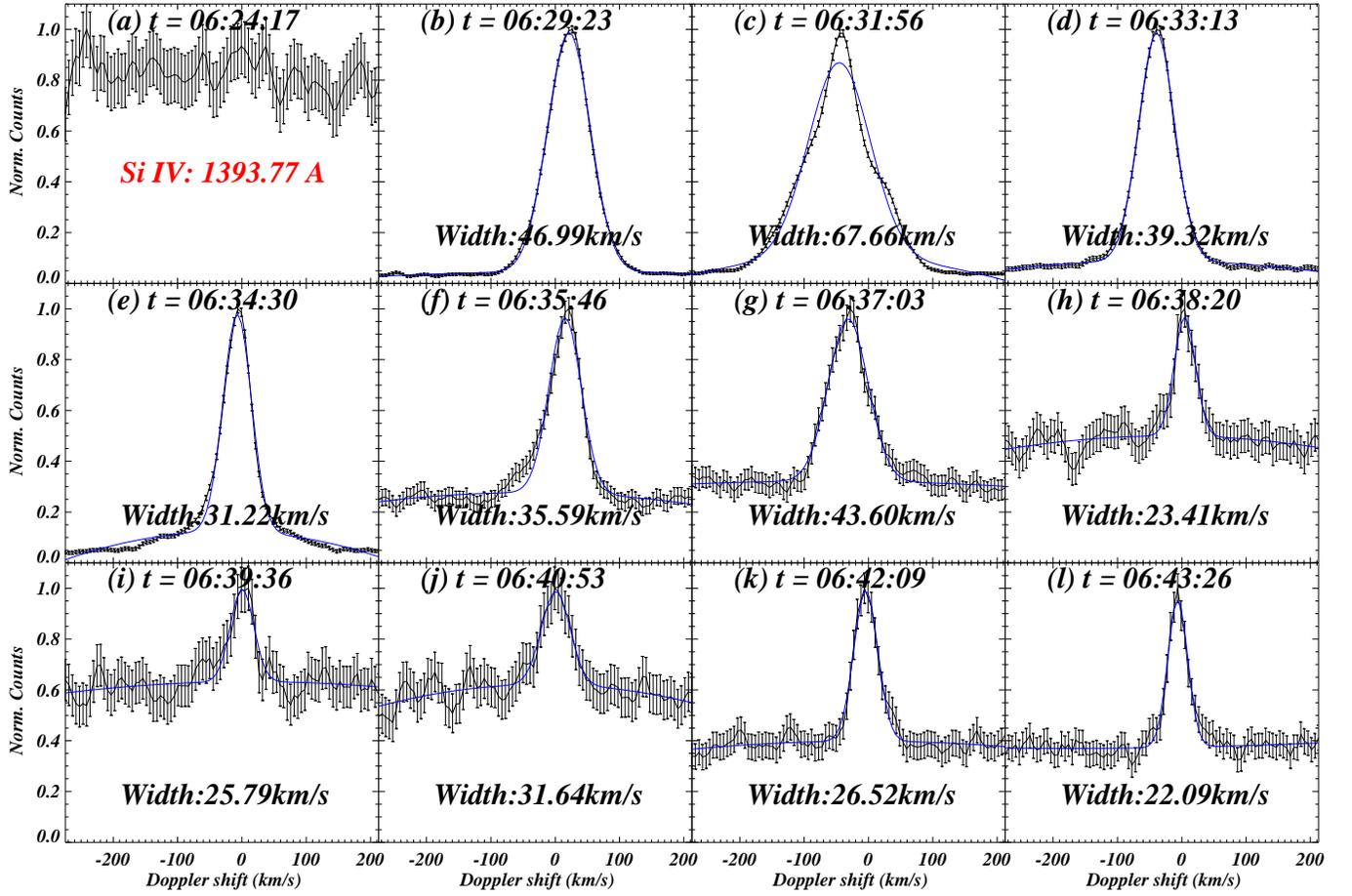}
  }
\caption{The temporal evolution of normalized averaged Si-IV line profiles (i.e., averaged over the five pixels around the blue plus location shown in Figure~\ref{fig:sji_spectra}) from the most probable reconnection region. It should be noted that all spectral profiles are normalized by their maximum
counts. All the spectral profiles are fitted by single Gaussian (see blue curve in all panels). We do see periodic fluctuations in the line width of Si~{\sc iv}. The first panel does not show the line as it is before the triggering of the solar jets.}
\label{fig:mr_evol}
\end{figure*}

Further, we have taken five pixels around blue plus location (see; panel (a) Figure~\ref{fig:sji_spectra}), and then all five profiles were averaged to get a single averaged Si~{\sc iv} profile at any particular instant of time. Through this approach, we have produced 33 averaged spectral profiles in a time range from 06:21:43~UT to 07:06:25~UT. This observation is an 8-step coarse raster observation with the cadence of 77 seconds. Therefore, the averaged spectral profile of any location is available after every 77 seconds (i.e., 8.0*cadence time). 

Some key averaged spectral profiles are displayed in the Figure~\ref{fig:mr_evol}. Panel (a) does not show the presence of the Si~{\sc iv} line as it is well before the jet event (t = 06:24:17~UT). While, t = 06:29:23 UT, we see the Si IV spectral line as the formation of the jet has already begun (see panel (b); Figure~\ref{fig:mr_evol}). We have fitted the line profile with the single Gaussian (blue curve) to estimate the peak intensity, Doppler velocity, and line width of the profile. The line width of this profile is high (i.e., 46.99 km/s) at time t = 06:29:23~UT (panel (b)). At next time t = 06:31:56~UT, we found that the profile is very wide and asymmetric too. We have fitted this profile with the single Gaussian, and found a very high line width (i.e., 67.66 km/s). It should be noted that the line width has increased a lot (i.e., 67.66 km/s) in comparison to the previous time t = 06:29:29~UT (panel (b). After t = 06:31:56~UT, we notice a decrease pattern in the line width of Si~{\sc iv} 1393.77~{\AA} profiles, i.e., the line width is 39.32 km/s at time t = 06:33:13 (panel (d) of Figure~\ref{fig:mr_evol}) and 31.22 km/s at time t = 06:34:30~UT (panel (e); Figure~\ref{fig:mr_evol}). Hence, for approximately 03 minutes (i.e., from 06:31:56~UT to 06:34:40~UT), the line width shows a decreasing pattern as the line width falls from 67.66 km/s to 31.22 km/s.

However, this decrease pattern in the line width breaks at time t = 06:35:46~UT as we see that the line width is now increasing with time (see; panels (f) and (g) of Figure~\ref{fig:mr_evol}). But again, the line width decreases with time t  = 06:38:20~UT (panel (h); Figure~\ref{fig:mr_evol}). And one more time (i.e., third time), we see the same behavior of the line width, i.e., the line width increase (panels (i) and (j); Figure~\ref{fig:mr_evol}) and decrease further with time (panels (k) and (l) of Figure~\ref{fig:mr_evol}). This particular finding indicates that the line width of the Si~{\sc iv} spectral line has periodic behavior.\\ 
 \begin{figure*}
  \mbox{
  \includegraphics[trim = 0.0cm 0.0cm 3.0cm 1.0cm, scale=1.0]{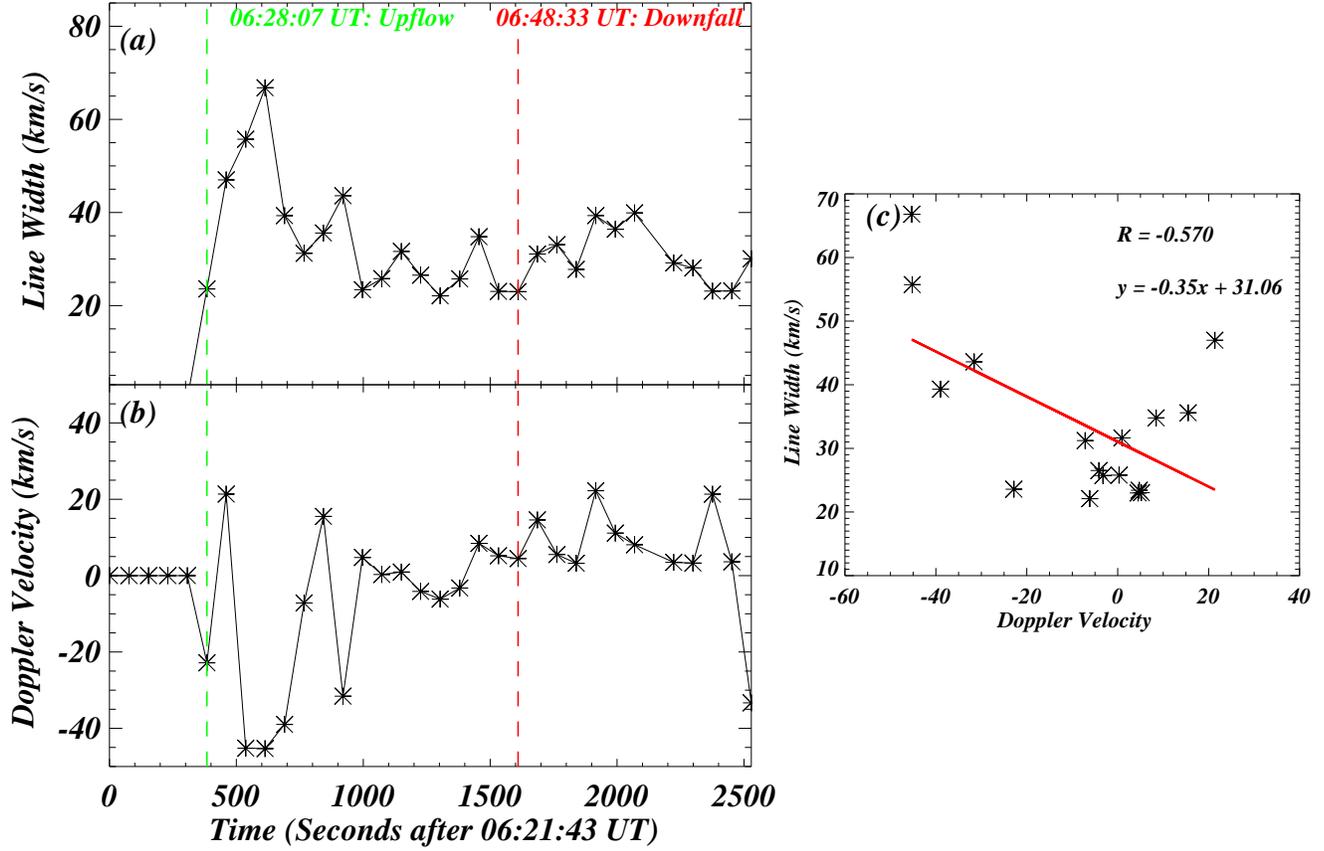}}
\caption{The temporal evolution of line width and Doppler velocity from the most probable reconnection region (blue plus sign in Figure~\ref{fig:sji_spectra}) is displayed in panels a and b, respectively. The triggering time of the jet is indicated by the green-dashed line, therefore, all the points before the green-dashed line are zero. The downfall phase of the blowout jet dominates after the red vertical dashed line. In the up flow phase of the jet, we do see the periodic behavior of the line width. Further, we also see the periodic behavior of Doppler velocity in the up-flow phase of the blowout jet (panel b). We have performed a correlation between line width and Doppler velocity which is shown in panel (c). It is found that line width is negatively correlated with the Doppler velocity, i.e., line width is during the up-flow, and the line width decreases as line profiles move towards the red-shifts.}
\label{fig:width_dv_qpp}
\end{figure*}

To understand the periodic behavior of line width clearly, we have plotted the line width with time (see; panel (a) of Figure~\ref{fig:width_dv_qpp}). As we know already, in the specified time range (i.e., t = 06:21:28~UT to 07:03:52~UT), there are 33 spectral lines, i.e., 33 line widths values. The specified time range starts from 06:21:28~UT, however, the jet appears in the selected region around 06:28:07~UT (see dashed green vertical line in panel (a); Figure~\ref{fig:width_dv_qpp}). We know that there is no spectral line before 06:28:07~UT, therefore, all the points before the green dashed vertical line have zero line width. We see the increase and decrease in the line width on a regular interval of time (see; panel (a); Figure~\ref{fig:width_dv_qpp}) . The oscillating behavior of the line width exists up to the time of t = 06:48~UT, i.e., up to the red-dashed vertical line. After $\sim$06:48~UT, the downfall phase of the blowout jet dominates, and we don't see much variations in the line width. So, finally, we can say that oscillations in the line width are present during the up-flow phase of the blowout jet.\\

In addition to the line width, we have also shown the Doppler velocity with the time in panel (b) of Figure~\ref{fig:width_dv_qpp}. The first few points are at zero Doppler velocity (up to the green dashed line) as the jet was not triggered by that time. And after that, we do see the fluctuation in the Doppler velocities (see points after the green dashed line). The careful inspection reveals that Doppler velocity is anti-correlated with line width during the up flow phase of the jet (i.e., points in between the green and red dashed lines). It means when the line width is high then Si~{\sc iv} line is blue-shifted and vice versa. As the line width decreases with time, then in response, the Si~{\sc iv} line moves towards the red shifts. We have already pointed out that there are a few cycles of periodic increase and decrease in the line width during the up-flow phase of the blowout jet. Similarly, we do see a kind of periodic behavior of Doppler velocity too. After 06:48:33~UT, all spectral profiles are red-shifted (see, points after the red dashed vertical line) as this is downfall dominant phase of the blowout jet. Further, we have checked the correlation between the line width and Doppler velocity for the up-flow dominated phase of the blowout jet (i.e., points between green and red dashed vertical lines) which is shown in the panel (c) of Figure~\ref{fig:width_dv_qpp}. Now, it is well clear that line width and Doppler velocity are anti-correlated, i.e., the blue shifts (upflows) have high line width, and when line width moves towards the red shifts (downflows) the line width decrease. The Pearson coefficient is quite good (i.e., -0.57) for this correlation. Hence, we can say that Doppler velocity and line widths are anti-correlated during the up-flow phase of the jet. 

\subsection{Quasi-periodic Pulsation}\label{sect:QPP}
We have found the presence of QPPs during the blowout jet. We have utilized AIA~304~{\AA}, AIA~171~{\AA}, AIA~131~{\AA}, and AIA~211{\AA} filters to study the QPPs in this blowout jet. 
\begin{figure*}
\mbox{
\centering
  \includegraphics[trim = -6.0cm -0.45cm -2.0cm 0.0cm, scale=0.9]{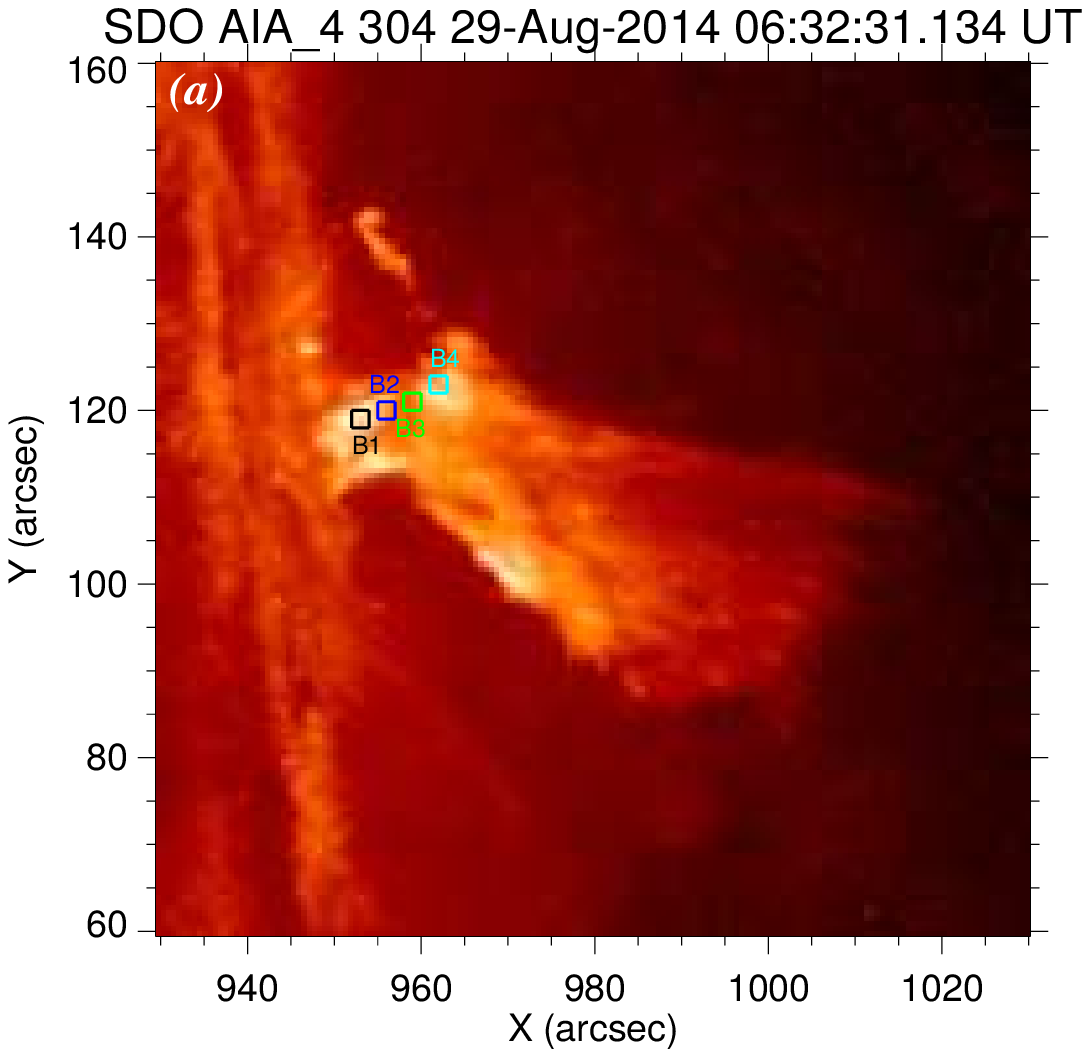}
}
\mbox{
  \includegraphics[trim = -1.0cm -0.45cm 0.0cm 0.0cm, scale=1.]{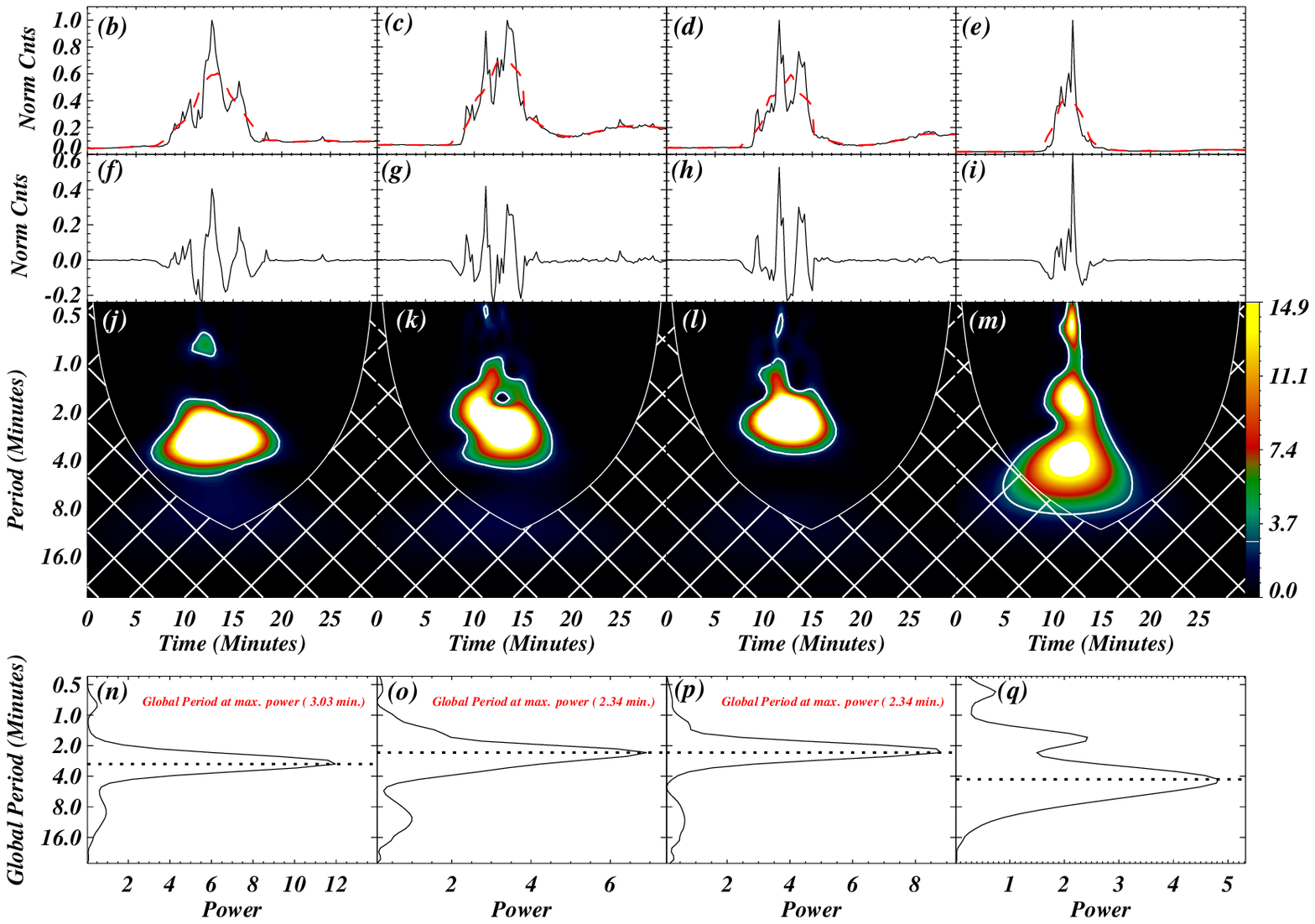}}
\caption{The panel (a) displays SDO/AIA 304 {\AA} image of the blowout jet at time t = 06:32:31~UT, and we select five boxes (B1, B2, B3, and B4) to deduce the emission curve from this filter. The temporal evolution of the intensity (i.e., black curve) in AIA 304 {\AA} filter from box B1 is displayed in the first panel of the first column. The over-plotted red dashed line is smoothed curve with a window of 15 points. The second panel of the first column shows the detrended curve, and the wavelet transform is applied to this detrended curve. The wavelet power map is displayed in the third panel with 95\% significance (i.e., white contours). The power is mainly concentrated around $\sim$03 minutes. Finally, in the last panel of first column, the global wavelet power is displayed which again shows that global power peaks around $\sim$ 03 minutes (i.e., 2.78 minutes). A similar analysis is shown for B2 (second column), B3 (third column), and B4 (fourth column), and the dominant period is $\sim$ 03 minutes. While we did not find any significant period in B4.}
\label{fig:wavelet_qpp}
\end{figure*}
We have selected four different boxes (i.e., B1, B2, B3, and B4) that are shown in the panel (a) of Figure~\ref{fig:wavelet_qpp} by different colors. Further, we estimated the averaged intensity curve (i.e., averaged over all pixels in the box) from all four boxes in all AIA filters (i.e, AIA~304~{\AA}, AIA~171~{\AA}, AIA~131~{\AA}, and AIA~211{\AA}). The panels (b) to (q) of Figure~\ref{fig:wavelet_qpp} show the wavelet analysis from four boxes. The wavelet analysis is performed on the averaged intensity light curves deduced from AIA~304~{\AA}. The panel (b) of Figure~\ref{fig:wavelet_qpp} shows the averaged AIA~304~{\AA} intensity curve (black curve) deduced from box B1 (black box in panel (a) of Figure~\ref{fig:wavelet_qpp}). There is an over-plotted red curve that is a smoothed averaged intensity curve with a window of 15 points. Further, the panel (f) shows the detrended intensity curve, i.e., averaged AIA~304~{\AA} light curve (black curve) - smoothed AIA~304~{\AA} curve (red curve; panel (b); Figure~\ref{fig:wavelet_qpp}). Then, we applied the wavelet analysis on this detrended light curve, and the deduced wavelet power map is shown in the panel (j) of Figure~\ref{fig:wavelet_qpp}. Here, we do see a concentrated patch of power around the period of 03 minutes (i.e., 3.03 minutes) for a time range from 06 to 20 minutes. Further, we have estimated a 95\% significance level that is important to check the reliability of any detected period in the wavelet analysis. And, the 95\% significance level is shown by a white contour on the wavelet power map. Now, it is visible that a concentrated patch of the power lies within the 95\% significance level contours. The cross-hatched gray area in the panel (j) of Figure~\ref{fig:wavelet_qpp} outlines the cone-of-influence (COI), and the powers inside this cross-hatched area are not reliable. But, here we can see that all the significant powers are outside of the COI. In the panel (n) of Figure~\ref{fig:wavelet_qpp}, we have shown the global power (i.e., wavelet power averaged over time) against the period. The global power shows dominant peak at a period of $\sim$03 minutes, i.e., 3.03 minutes.\\ 

We have applied the wavelet analysis in the same fashion to all other boxes (i.e., B2, B3, and B4) which are shown in panel (a) of Figure~\ref{fig:wavelet_qpp}. The original $\&$ smoothed intensity curves, detrended intensity curve, wavelet power maps, and global power maps are shown in the same manner for box B2 (panels (c), (g), (k), and (o)), box B3 (panels (d), (h), (l), and (p)), and box B4 (panels (e), (i), (m), and (q)) in the Figure~\ref{fig:wavelet_qpp}. Boxes B2 and B3 show very similar behavior as we found for box B1. The dominant period is also approximately 03 minutes (i.e., 2.34 minutes for box B2 (panel (o) of Figure~\ref{fig:wavelet_qpp}) and box B3 (panel (p) of Figure~\ref{fig:wavelet_qpp})) for the almost same time range from 06 to 20 minutes. \\ 

However, the intensity curve of the last box (i.e., B4) shows a sharp jump in the intensity for a short interval of time (i.e., around 05 minutes only). It is unlike to the other boxes (B1, B2, and B3) as fluctuations sustain a bit longer therein (around 15 minutes). We applied wavelet analysis in the same manner to box B4 also, and we found three concentrated patches of the power around the period of 06, 03, and 0.5 minutes (panel (q) of Figure~\ref{fig:wavelet_qpp}). The wavelet power patches around the period of 06, 03, and 0.5 minutes persist only for 10 minutes, 05 minutes, and less than one minute, respectively. Hence, these periods of box B4 (i.e., 06, 03, and 0.5 minutes; panel (q) of Figure~\ref{fig:wavelet_qpp}) do not even complete two cycles, therefore, we are assuming them as non-reliable power. In addition, some part of the longer period (06 minutes) also lies within the COI (panel (m) of Figure~\ref{fig:wavelet_qpp}). Hence, finally, we can say that none of the power patches in this wavelet power map of box B4 is reliable. And, we mention that the B4 does not has any periodicity unlike the other boxes (i.e., B1, B2, and B3). We have also shown the QPPs in the same fashion for AIA~171{\AA} (cf., Figure~\ref{fig:wavlet_qpp_171}), AIA~131{\AA} (cf., Figure~\ref{fig:wavelet_qpp_131}), and AIA~211{\AA} (cf., Figure~\ref{fig:wavelet_qpp_211}) in the appendix~\ref{sect:append1}. The findings from these fitters are similar to what we have reported for AIA~304{\AA} here. 
\section{Discussion and Conclusions} \label{sect:discuss}
The present work provides an observation of the formation of blowout jet through kink instability. Initially, an inverse $\gamma$-shape flux-rope appears on the west limb on August 29$^{th}$, 2014 that is a morphological indication for the onset of kink instability \citep{2005ApJ...630L..97T, 2009ApJ...691...61P, 2013ApJ...770L...3K, 2016ApJ...832..106H}. The inverse $\gamma$-shape flux-rope activates around 06:28:00~UT, i.e., this structure rises, and expands with time. The twisted field lines are associated with inverse $\gamma$-shape flux-rope, and these magnetic field lines reconnect. The primary magnetic reconnection takes place around 06:31:00~UT near the apex of the inverse $\gamma$-shape flux-rope, i.e., in the vicinity of the bright knot. We have witnessed the bi-directional flows from the apex of the flux rope through the imaging analysis (section~\ref{sect:multi_jet}). Various spectral lines (i.e., Si~{\sc iv}, C~{\sc ii}, and Mg~{\sc ii}) clearly show red-shifted profiles (i.e., plasma downfall) below the apex, and all these profiles become blue-shifted (i.e., plasma up flow) above the apex of flux-rope (cf., Figure~\ref{fig:sji_spectra}), i.e., bi-directional flows. Hence, both (images and spectra) confirm the presence of bidirectional flows which is a typical characteristic of the magnetic reconnection in the solar atmosphere \citep{1997Natur.386..811I, 2014ApJ...797...88H, 2015ApJ...813...86I, 2020ApJ...898..101Y, 2020ApJ...890L...2C, 2021NatAs...5..237B, 2021SoPh..296...84D, 2021NatAs...5...54A}. Further, DEM analysis shows the presence of multi-thermal plasma around the knot of inverse $\gamma$-shape flux-rope in the wide temperature range (log T/K = 5.4{--}7.2; Figure~\ref{fig:dem_jet}). Hence, these observational findings (i.e., bi-directional flows and multi-thermal plasma) indicate that the magnetic reconnection (primary) takes place around the knot of inverse $\gamma$-shape flux-rope which triggers the jet. \\ 
Soon after the primary magnetic reconnection, the northern leg of the inverse $\gamma$-shape flux-rope completely erupts, and further, the jet has developed along only the southern leg. Interestingly, we have seen the multiple bright regions (with time) within the jet. Here, we have clearly seen the multiple bright spikes in the time-distance diagram estimated as per the slit (i.e., S1) along the blowout jet. Similarly, the time-distance diagrams estimated as per the slits (i.e., P2, P3, and P3) across the blowout jet show the multiple bright dots (cf.,(section~\ref{sect:td})). It is trivial to understand that these multiple spikes (along the jet) or multiple bright dots (across the jet) are forming due to multiple enhancement in the intensity with time. Most probably, this multiple enhancements in the intensity supports the multiple magnetic reconnection scenario \citep{2012A&A...542A..70M, 2015ApJ...807...72L, 2017ApJ...836..121K}. Spectroscopic observations reveal a periodic enhancement in the line width of Si~{\sc iv} 1393.77~{\AA} (cf., Figure~\ref{fig:mr_evol} and~\ref{fig:width_dv_qpp}). For the first time, the periodic enhancement of Si~{\sc iv} line width is being reported in this blowout jet event. On top of these crucial observational findings, we have seen that Si~{\sc iv} profiles are blue-shifted (upflows) when they are very broad (i.e., high line width). And, gradually, the profiles become narrower while they are moving toward the red-shifts (down flows). The periodic existence of such broadened blue-shifted Si~{\sc iv} profiles is most probably due to the occurrence of multiple magnetic reconnection in this blowout jet. Most importantly, our observations also reveal very complex and explosive type profiles of some prominent spectral lines (i.e., Si~{\sc iv}, C~{\sc ii}, and Mg~{\sc ii}) of the solar interface-region (cf., Figure~\ref{fig:sji_spectra}). As we know that such complex $\&$ explosive type profiles are produced only due to the magnetic reconnection \citep{2014Sci...346C.315P, 2015ApJ...813...86I, 2017MNRAS.464.1753H, 2018ApJ...857....5Y, 2020ApJ...890L...2C}, all the observational findings indicate the occurrence of multiple magnetic reconnection. \\

QPPs in the solar/stellar flares is an often phenomenon that occurs with few seconds to few minutes of oscillations period. Several reports discuss the triggering and related dynamics of QPPs in the solar atmosphere \citep[e.g., ][]{2006SoPh..238..313C, 2009A&A...493..259I, 2009SSRv..149..119N, 2016SoPh..291.3143V, 2017A&A...597L...4L, 2018ApJ...859..154N, 2018SSRv..214...45M, 2020A&A...642A.195K, 2022Univ....8..104S, 2022ApJ...930L...5Z}. The statistical studies of the intense solar flares suggest that the occurrence rate of QPPs reaches 30-80 $\%$ with intense flares lying above the M5 class \citep{2015SoPh..290.3625S}. However, the QPPs occurrence rate reduces with the low intense solar flares \citep{2021SSRv..217...66Z}. Using GOES X-ray data from 2011 to 2018, \citep{2020ApJ...895...50H} performed the statistical analysis for QPPs and their association with the different classes of solar flares. The authors claimed that the $\approx$46$\%$ of X-class and $\approx$29$\%$ of M-class flares show QPPs signature. However,  only $\approx$9$\%$ of C-class flares exhibit QPPs signature. On the other hand, there are few reported observations of QPPs in jets \citep{2012A&A...542A..70M, 2014A&A...561A.134Z, 2018MNRAS.480L..63S}. Interestingly, in the present work, the wavelet analysis of light curves from five different boxes in various AIA filters clearly demonstrates the existence of QPPs in the present blowout jet (section~\ref{sect:QPP}).\\

The triggering mechanisms of QPPs are very crucial, and so far, more than 15 mechanisms have been proposed to understand the initiation mechanism of the QPPs \citep{2009SSRv..149..119N, 2016SoPh..291.3143V, 2021SSRv..217...66Z}. Broadly, these triggering mechanisms of QPP may be classified into two categories, namely, periodic spontaneous magnetic reconnection \citep{2000A&A...360..715K, 2005A&A...432..705K, 2009A&A...494..329M, 2012A&A...542A..70M, 2015ApJ...807...72L, 2018SSRv..214...45M}  and the MHD waves that may induce the periodic magnetic reconnection \citep[e.g.,][]{2004A&A...419.1141N,2005A&A...440L..59F, 2006A&A...446.1151N, 2011ApJ...730L..27N, 2016ApJ...823L..16T, 2016ApJ...832...65Z, 2016ApJ...822....7K, 2021SSRv..217...66Z}. \citet{2004A&A...419.1141N} found recurring explosive events with a period of 3{--}5 minutes when the compressible waves push the oppositely directed field lines to reconnect. This process leads to multiple magnetic reconnection, and the QPPs were triggered by multiple magnetic reconnection (i.e., recurring explosive events). The other wave modes (e.g., fast mode MHD waves and global kink mode) may also trigger the periodic magnetic reconnection in a coronal loop that is situated near the flaring region. It initiates QPPs with a period of several minutes (e.g., \cite{2005A&A...440L..59F, 2006A&A...446.1151N}). In the present work, the detected QPPs from various intensity light curves in the blowout jet have a period of $\sim$ 03 minutes (section~\ref{sect:QPP}). The extracted EM from the blowout jet in the temperature range of 0.5{--}15 MK also shows the temporal variations on a time-scale of $\sim$ 03 minutes (section~\ref{sect:dem}; Figure~\ref{fig:dem_jet}). Apart from the intensity and EM curves, the line-width of Si~{\sc iv} is also fluctuating on a time scale of approximately 03 minutes (section~\ref{sect:spectra}). Hence, consistently, we have found the fluctuations at a time scale of $\sim$ 03 minutes in various parameters (e.g., intensity, EM, and line width). Here, we would like to mention that this blowout jet triggers within an active region (AR). The umbra of the sunspot (i.e., photospheric and chromospheric atmosphere of AR) is filled with the 3-minute slow MHD waves (e.g.,\citealt{1991A&A...250..235F, 2014ApJ...786..137T, 2012ApJ...757..160J, 2017ApJ...836...18C,2019A&A...627A.169F,2021NatAs...5....2F, 2011ApJ...728...84B, 2020ApJ...903...19F, 2021ApJ...906..121K}). Hence, we conjecture that 03-minute oscillations are present within the triggering site of the blowout jet, and they may drive the periodic magnetic reconnection at time-scale of $\sim$ 03 minutes. Hence, most probably, we can say that the periodic magnetic reconnection produces the observed periodic fluctuations (i.e., QPPs) in the intensity, EM, and line width.\\ 

It should be noted that the magnetic reconnection between pre-existing and open coronal field and closed magnetic fields can produce a collimated jet without the rotation or twist, i.e., a kind of standard jet \citep[e.g.,][]{1996PASJ...48..353Y, 2003ApJ...593L.133M,  2008ApJ...673L.211M, 2010ApJ...720..757M, 2011ApJ...735L..18L, 2013ApJ...771...20M, 2016SSRv..201....1R, 2021RSPSA.47700217S}. On the other hand, the magnetic reconnection between the pre-existing open coronal magnetic field and the twisted closed magnetic fields can produce newly twisted magnetic field lines which undergo the untwisting motions (e.g., \citealt{2014ApJ...789L..19F}). The plasma flows along the newly twisted magnetic field lines that form the solar jet, and the rotational/helical motion of the solar jets is a result of untwisting motion of these newly twisted magnetic field lines. Hence, the magnetic reconnection is an important physical process to trigger the solar jets in the solar atmosphere \citep{1992PASJ...44L.173S, 1995Natur.375...42Y, 1996PASJ...48..123S, 1998SoPh..178..379S, 2007Sci...318.1591S, 2013ApJ...770L...3K, 2015Natur.523..437S, 2014Sci...346A.315T, 2015A&A...581A.131J, 2017Natur.544..452W, 2018A&A...616A..99K, 2021ApJ...920...18S}. That is why this physical process (i.e., magnetic reconnection) is an integral feature of various 2.5D and 3D models of the solar jets \citep{1996PASJ...48..353Y, 2008ApJ...683L..83N, 2008ApJ...673L.211M, 2009A&A...506L..45G, 2015A&A...573A.130P, 2016A&A...596A..36P}. The present blowout jet shows a very strong rotation of the plasma column (see attached animation$\_$1.mp4). 
The helical or rotational motions of the solar jets are an important indication of the kink instability (e.g., \citealt{1996AdSpR..17d.197S, 2016SSRv..201....1R, 2021RSPSA.47700217S}). The numerical simulations have shown that destabilization of the system by global kink-instability (when helicity or twist exceeded the critical value) can trigger the magnetic reconnection through the separatrix surface, and the magnetic reconnection drives the helical solar jets (e.g., \citealt{2009ApJ...691...61P, 2010ApJ...714.1762P, 2010ApJ...715.1556R,2015A&A...573A.130P}).
Most of the magnetic field is either open or long curvature magnetic field in the vicinity of the blowout jet (see panels (c1), and (d1) of Figure~\ref{fig:jet_trigger}). Here, we mention that kink instability destabilizes the pre-existing coronal magnetic field configuration through either magnetic reconnection between the kinked flux rope (i.e., kinked/erupting loops) and the pre-existing coronal fields or the internal magnetic reconnection within the kink unstable flux rope. Hence, due to the magnetic reconnection, the plasma flows along reconnected magnetic field lines which collectively form the spire of the blowout jet. This blowout jet is mainly visible in the cool-temperature filters, and the signature of this blowout jet is faint in the hot-temperature filters (see section~\ref{sect:multi_jet}). However, usually, the blowout jets have strong emissions at hot temperatures along with strong emissions at cool temperatures (e.g.,\citealt{2010ApJ...720..757M,2013ApJ...769..134M}). Here, it should be noted that surges emit mainly at the cool temperature. Therefore, in the present observational baseline, we don't rule out the possibility of this jet being an chromospheric surge. However, the emission measure (EM) in high-temperature wavebands also shows some hot plasma emission (4-6 MK) from the spire of this jet. Therefore, this manuscript uses this feature as a blowout jet.

Hence, kink-instability is an possibly an important driver of solar jets, and not only in the solar jets, but the kink-instability also plays crucial role in the triggering of the large-scale eruptions of the solar atmosphere \citep{2004A&A...413L..27T, 2005ApJ...630L..97T, 2010ApJ...715..292S, 2013ApJ...765L..42S, 2012ApJ...746...67K,  2021NatCo..12.2734Z}. In case of observations related to the kinked solar flux-rope in the solar jets, \citet{2013ApJ...770L...3K} have reported a kinked flux-tube that drives a solar jet, i.e., internal magnetic reconnection in the kinked flux-tube at the north polar triggers the polar jet. Further, \citet{2017ApJ...844L..20Z} have also shown that kink instability triggers a blowout jet in the solar atmosphere. The present observation clearly shows the occurrence of the kinked flux rope at the west limb prior to the jet formation. Further, the imaging, as well as spectroscopic observations, confirms the multiple magnetic reconnection in support of the formation of this solar blowout jet.

Hence, in conclusion, we say that the present observational baseline shows the inverse $\gamma$-shape, rotational or helical motion, and multiple magnetic reconnection in this blowout jet event. Most probably, these observational findings collectively indicate that the kink-instability triggers this blowout jet, and multiple magnetic reconnection leads to the formation of QPPs in this jet.
\section{Acknowledgements}
S. K. Mishra acknowledges the Indian Institute of Astrophysics (IIA, Bangalore) for providing the computational facilities and institute fellowship. K. Sangal would like to acknowledge the Council of Scientiﬁc \& Industrial Research (CSIR), Government of India, for ﬁnancial support through a Senior Research Fellowship (UGC-SRF). P. Jel\'{i}nek acknowledges support from grant 21-16508J of the Grant Agency of the Czech Republic. A. K. Srivastava acknowledges the ISRO Project Grant (DS\_2B-13012(2)/26/2022-Sec.2) for the support of his research. S.P. Rajaguru acknowledges support from SERB (Govt of India) research grant CRG/2019/003786. We acknowledge the use of \citep{2012A&A...539A.146H} for calculating the differential emission measure (DEM). Data courtesy of SDO/AIA and IRIS science team. We also acknowledge the use of \citep{1998BAMS...79...61T} method to extract the wavelet power spectra and average period of QPPs.


\appendix
\section{Temporal evolution of jet in IRIS/SJI 1330~{\AA} filter} \label{sect:append2}
We have shown the evolution of the jet in IRIS/SJI~1330~{\AA} filter observations (cf., figure~\ref{fig:jet_evol_iris}). At time t = 06:29~UT, we have seen the activation of the flux rope at the west limb of the Sun (panel a). Further, we have seen the formation of the bright knots that are most probably forming due to the magnetic reconnection (panel d), and this magnetic reconnection triggers the jet (panel d). The magnetic reconnection happens in the vicinity of bright knots, therefore, below the bright knots, the plasma falls back towards the limb (downflows) along both legs (indicated by red arrows in panel f). While above the bright knot the plasma flows upward as indicated by the blue arrow in panel (f). Around t = 06:33~UT, the northern leg of the jet erupts completely (panels h), and further, the jet evolves around the southern leg (panels h and i). In the later phase of the jet, we have seen that most of the plasma falls back towards the solar surface. The evolution and dynamics of the jet in this filter (i.e., IRIS/SJI~1330~{\AA}) are very similar as we have already seen in the AIA~304~{\AA} filter (cf., figure~\ref{fig:kink_unstable}). The dynamics and evolution of this jet with the help of AIA~304~{\AA} filter in the main text. Similar to AIA~304~{\AA} filter, we have tracked a plasma thread that shows rotation with time (see curved white arrows from panels d to f). A similar evolution of this jet can be seen in the given animations (i.e., animation$\_$1.mp4{--}with annotation and animation$\_$2.mp4{--}without annotation). The real-time duration of IRIS animations is 23 seconds. Here, it should be noted that IRIS observations are rotated by a roll angle of $90^\circ$, and we have provided the direction arrows in the first panel of figure~\ref{fig:jet_evol_iris}.
\setcounter{figure}{0}
\renewcommand\thefigure{\thesection.\arabic{figure}}
\begin{figure*}
    \centering
    \includegraphics{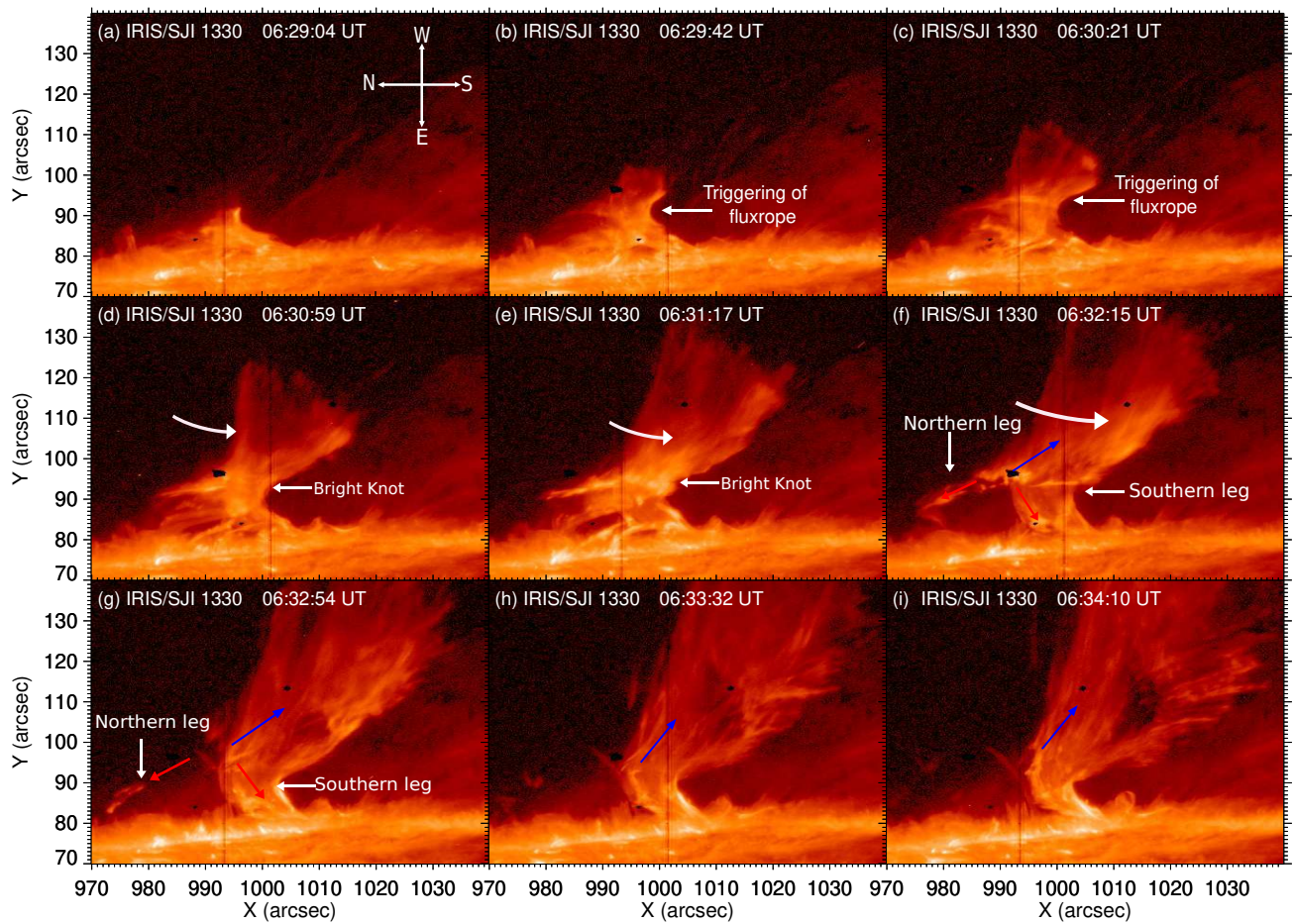}
    \caption{This figure shows the evolution of the jet in the IRIS/SJI 1330~{\AA} filter. Firstly, we saw the activation of the kinked flux rope (panels b and c), and then the formation of bright knots due to the magnetic reconnection of twisted field lines (panel d). This magnetic reconnection leads downflows along both legs of the jet (see red arrows in panel f) and upflows along the spire of the jet (blue arrow in panel f). Further, after some time, one leg erupts completely (panels g and h), and the jet further develops along the southern leg (panels h and i). We also see the rotation of the plasma as indicated by white arrows from panel (d) to panel (f). At last, we mention that evolution of the jet in IRIS/SJI~1330~{\AA} is similar as we have already seen in AIA~304~{\AA} (cf., figure~\ref{fig:kink_unstable}). IRIS animations (i.e., animation$\_$1.mp4{--}with annotation and animation$\_$2.mp4{--}without annotation) also show the same evolution of this jet. The IRIS animations start from 06:24:45~UT to 07:00:50~UT having a real time duration of 23 seconds.}
\label{fig:jet_evol_iris}
\end{figure*}
\section{Quasi Periodic Pulsations: AIA~171~{\AA}, AIA~211~{\AA}, and AIA~131~{\AA}} \label{sect:append1}
In this appendix, we have shown the wavelet analysis for AIA~171~{\AA} (Figure~\ref{fig:wavlet_qpp_171}), AIA~131~{\AA} (Figure~\ref{fig:wavelet_qpp_131}), and AIA~211~{\AA} (Figure~\ref{fig:wavelet_qpp_211}). We have shown wavelet power maps from all four boxes in the same fashion as we have described in section~\ref{sect:QPP}. Interestingly, the QPPs in these hot EUV wavebands of SDO/AIA also show a similar period as we found for AIA~304~{\AA} (section~\ref{sect:QPP}). 
\setcounter{figure}{0}
\renewcommand\thefigure{\thesection.\arabic{figure}}
\begin{figure*}
    \centering
    \includegraphics{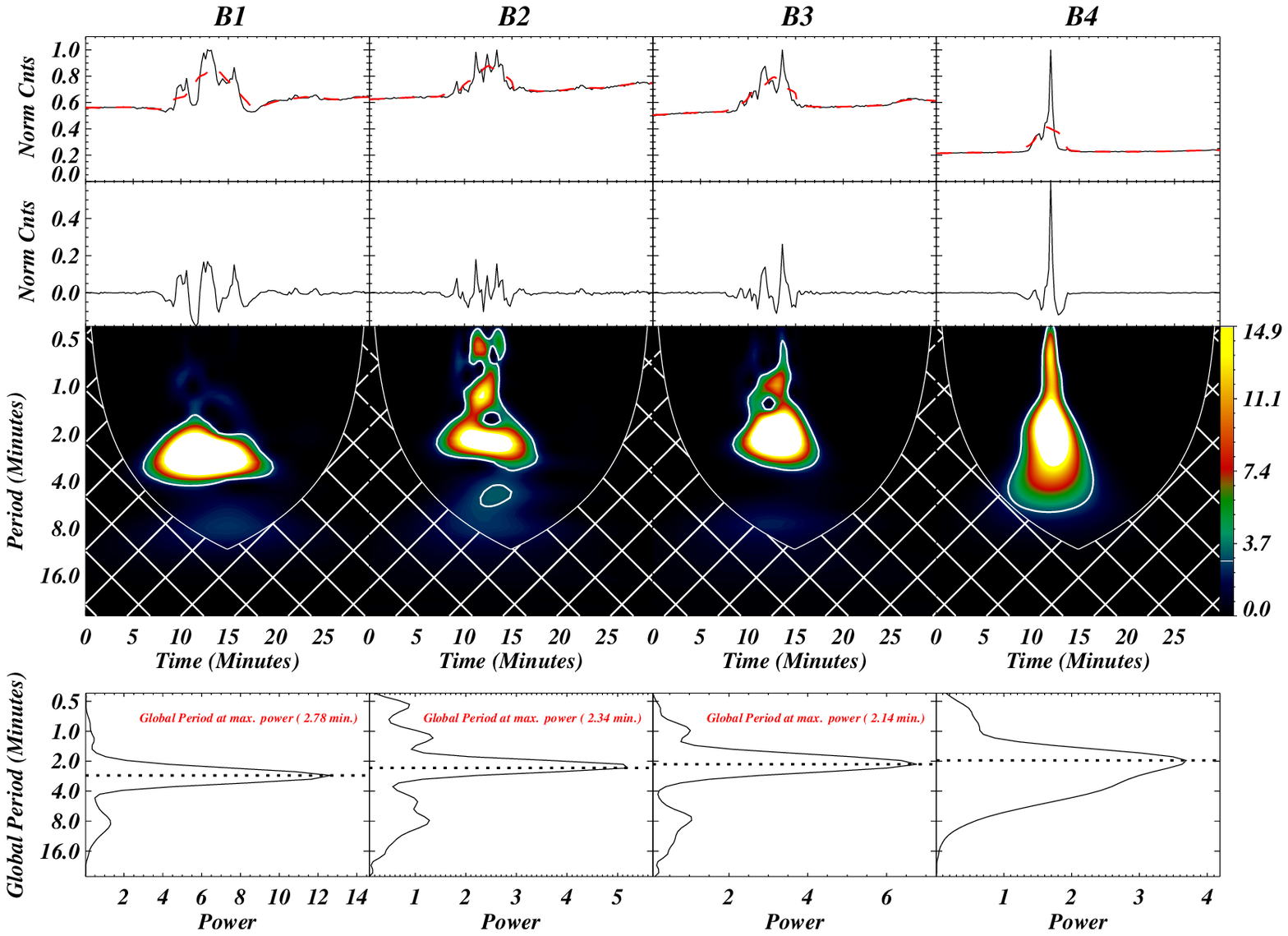}
    \caption{Same as figure~\ref{fig:wavelet_qpp} but for AIA~171~{\AA} filter observations.}
    \label{fig:wavlet_qpp_171}
\end{figure*}
\begin{figure*}
    \centering
    \includegraphics{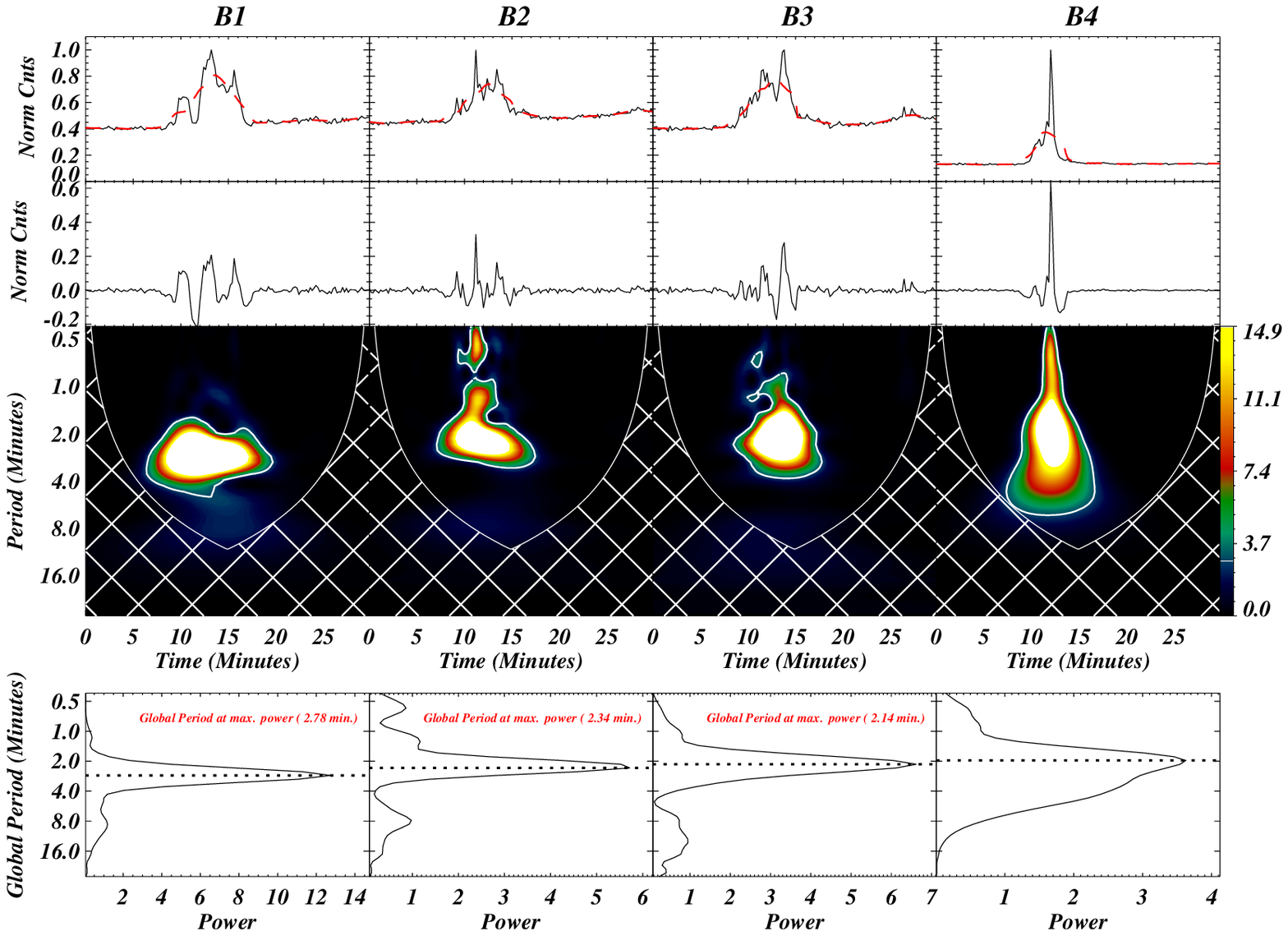}
    \caption{Same as figure~\ref{fig:wavelet_qpp} but for AIA~131~{\AA} filter observations.}
    \label{fig:wavelet_qpp_131}
\end{figure*}
\begin{figure*}
    \centering
    \includegraphics{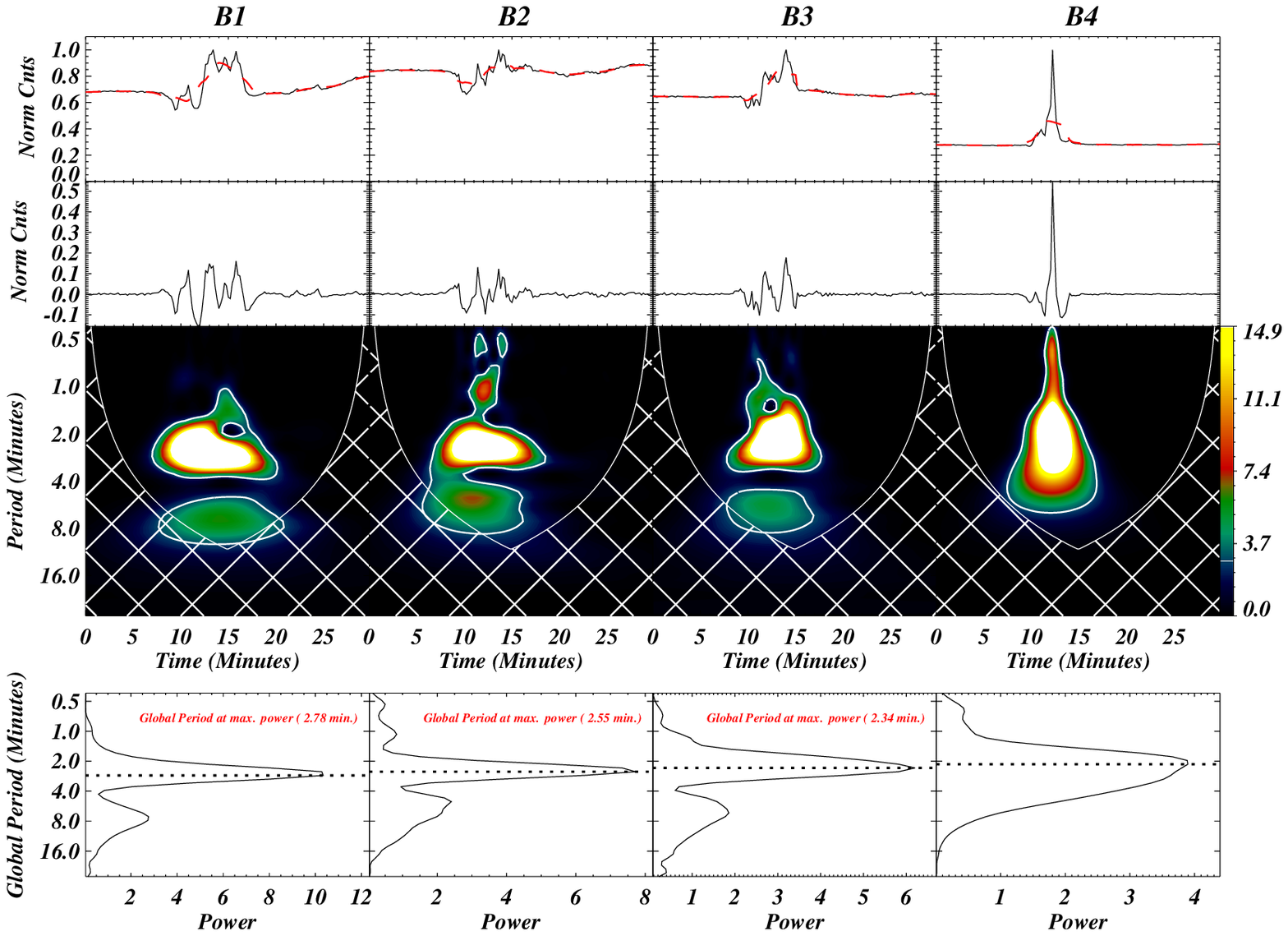}
    \caption{Same as Figure~\ref{fig:wavelet_qpp} but for AIA~211~{\AA} filter observations.}
    \label{fig:wavelet_qpp_211}
\end{figure*}
\end{document}